\documentclass{aastex}
\usepackage{emulateapj5}
\begin{document}

\newcommand{\hi}{$h^{-1}$~}
\newcommand{\kms}{~km~s$^{-1}$}
\newcommand{\logh}{$+5\log h$}

\title{A Keck Spectroscopic Survey of MS~1054--03 ($z=0.83$):\\
  Forming the Red Sequence\altaffilmark{1,2}}

\author{Kim-Vy H. Tran\altaffilmark{3,4,5,6}, Marijn Franx\altaffilmark{6}, Garth
  D. Illingworth\altaffilmark{7}, \\ 
  Pieter van Dokkum\altaffilmark{8}, Daniel D. Kelson\altaffilmark{9}, John
  P. Blakeslee\altaffilmark{10}, and Marc Postman\altaffilmark{11}}

\altaffiltext{1}{Based on observations obtained at the W. M. Keck 
Observatory, which is operated jointly by Caltech and the University
of California.} 
\altaffiltext{2}{Based on observations with the NASA/ESA Hubble Space
Telescope, obtained at the Space Telescope Science Institute, which
is operated by the Association of Universities for Research in
Astronomy, Inc., under NASA contract NAS 5-26555.}
\altaffiltext{3}{NSF Astronomy \& Astrophysics Fellow}
\altaffiltext{4}{NOVA Fellow}
\altaffiltext{5}{Harvard-Smithsonian Center for Astrophysics, 60
  Garden Street, Cambridge, MA 02138}
\altaffiltext{6}{Leiden Observatory, Leiden University, Niels Bohrweg
  2, 2333 CA Leiden, The Netherlands} 
\altaffiltext{7}{University of California Observatories/Lick
Observatory, University of California, Santa Cruz, CA 95064}
\altaffiltext{8}{Department of Astronomy, Yale University, New Haven,
CT 06520-8101}
\altaffiltext{9}{Observatories of the Carnegie Institution of
Washington, 813 Santa Barbara Street, Pasadena, CA, 91101}
\altaffiltext{10}{Department of Physics and Astronomy, Washington
  State University, Pullman, WA 99164-2814} 
\altaffiltext{11}{Space Telescope Science Institute, 3700 San Martin
  Drive, Baltimore, MD 21218} 

\setcounter{footnote}{11}

\begin{abstract}

Using a magnitude-limited, spectroscopic survey of the X-ray luminous
galaxy cluster MS~1054--03, we isolate 153 cluster galaxies and
measure MS1054's redshift and velocity dispersion to be
$z=0.8307\pm0.0004$ and $\sigma_z=1156\pm82$\kms. The absorption-line,
post-starburst (``E+A''), and emission-line galaxies respectively make
up $63\pm7$\%, $15\pm4$\%, and $23\pm4$\% of the cluster population.
With photometry from HST/ACS, we find that the absorption-line members
define an exceptionally tight red sequence over a span of $\sim3.5$
magnitudes in $i_{775}$: their intrinsic scatter in
$(V_{606}-i_{775})$ color is only $0.048\pm0.008$, corresponding to a
$(U-B)_z$ scatter of 0.041.  Their color scatter is comparable to that
of the ellipticals ($\sigma_{Vi}=0.055\pm0.008$), but measurably smaller
than that of the combined E+S0 sample ($\sigma_{Vi}=0.072\pm0.010$).
The color scatter of MS1054's absorption-line population is
approximately twice that of the ellipticals in Coma; this difference
is consistent with passive evolution where most of the absorption-line
members ($>75$\%) formed by $z\sim2$, and all of them by $z\sim1.2$.
For red members, we find a trend ($>95$\% confidence) of weakening
H$\delta$ absorption with redder colors that we conclude is due to
age: in MS1054, the color scatter on the red sequence is driven by
differences in mean stellar age of up to $\sim1.5$ Gyr.  We also
generate composite spectra and estimate that the average S0 in MS1054
is $\sim0.5-1$ Gyr younger than the average elliptical; this
difference in mean stellar age is mainly due to a number of S0s that
are blue (18\%) and/or are post-starburst systems (21\%).

\end{abstract}

\keywords{galaxies: clusters: individual (MS 1054-03) -- galaxies:
  elliptical and lenticular, cD -- galaxies: fundamental parameters --
  galaxies: evolution }

\section{Introduction}

Understanding how galaxies form and evolve in clusters continues to be
a fundamental question in astronomy.  The ages and assembly histories
of galaxies in rich clusters test both stellar population models and
hierarchical formation scenarios.  Studies of the passive, red
galaxies that dominate local clusters show that the bulk of their
stars formed at $z>2$ \citep{bower:92,jorgensen:99,trager:00} and
indicate that the cluster population has not evolved strongly in the
last $\sim8$ Gyr.  However, the current paradigm of hierarchical
assembly \citep{peebles:70} predicts that clusters continue to grow by
accreting galaxies.  Indeed, observations of clusters at $z>0.3$ have
revealed an increasing fraction of blue, star-forming galaxies at
higher redshift \citep{butcher:84,couch:87,ellingson:01}.

How to link the evolving cluster populations observed at intermediate
redshifts ($0.3<z<1$) to the predominantly quiescent ones at $z\sim0$
is a matter of much debate.  Using imaging from the {\it Hubble Space
Telescope}, several studies find a deficit of S0 galaxies in
$z\gtrsim0.4$ clusters and argue that the observed excess of blue
galaxies must evolve into the missing S0s
\citep{dressler:97,lubin:02,tran:05a,postman:05}.  However, this
implies that S0s are younger than the ellipticals, and such a
difference in mean age is at odds with other HST studies of
$z\sim0.3-0.5$ clusters that find the S0s to be as old as the
ellipticals \citep{ellis:97,kelson:00c}.  It may be that the
fractional age difference between Es and S0s is negligible at
$z\lesssim0.4$, but that such an imprint is still visible at
$z\gtrsim0.8$ when even massive cluster galaxies are at most only
$\sim4$ Gyr old \citep{thomas:05}.

This debate raises the more general issue of how the color-magnitude
relation, and subsequently the red sequence, forms and evolves.  The
ellipticals and S0 galaxies in local clusters define a narrow ridge on
the color-magnitude diagram that indicates homogeneously old ages
\citep{sandage:78,bower:92,kodama:98}, and the low color scatter in
the E+S0 members has been confirmed to $z\sim0.9$ \citep{stanford:98}.
However, if the galaxy types differ in mean stellar age
\citep[$e.g.$][]{poggianti:01b}, this implies that the red sequence
continued to assemble after it was initially seeded at higher redshift
($z>2$).  For example, faint ($L<L^{\ast}$) S0s are inferred to evolve
from fading blue galaxies at $z\lesssim0.8$, thus they appear on the
red sequence at lower redshifts compared to the more massive S0s
\citep{tran:05a,holden:06}, and the resulting red sequence galaxies
would have heterogenous star formation histories
\citep{vandokkum:98a,bower:98}.  Therefore determining how the red
sequence assembles by measuring, $e.g.$ how the color scatter evolves,
provides useful tests of cluster formation models
\citep{kauffmann:95,baugh:96,diaferio:01}.

The main challenges for studying cluster populations at intermediate
redshifts and how they evolve are 1) isolating the members and 2)
obtaining the high resolution imaging needed to determine
morphologies.  Although photometric surveys have an advantage in terms
of surveyed area, they have inherent and possibly large uncertainties
due to field corrections.  Only with spectroscopy can we confidently
isolate cluster galaxies from the field.  Spectroscopy also provides
useful and independent diagnostics for separating members into
absorption-line, emission-line, and post-starburst
\citep[``E+A'';][]{dressler:83} galaxies, and for determining mean
stellar ages to identify, $e.g.$ the oldest cluster galaxies.

Motivated by these issues, we present an extensive magnitude-limited
spectroscopic survey of MS~1054--03, a massive, X-ray luminous galaxy
cluster first detected in the {\it Einstein} Medium
Sensitivity Survey \citep{gioia:94} and spectroscopically confirmed
to be at $z=0.83$ by \citet{donahue:98}.  With medium resolution
spectroscopy from Keck/LRIS \citep{oke:95}, we isolate 153 cluster
galaxies and measure their spectral indices.  We pair the spectroscopy
with imaging from the {\it Advanced Camera for Surveys}
\citep{ford:98} that provides accurate colors and magnitudes as well
as morphological types.  To improve our spectral signal-to-noise, we
also generate composite spectra for the different spectral and
morphological classes, and compare their spectral indices to stellar
population models.

Our primary goal is to trace how the red sequence assembles.  To do
so, we first identify the oldest cluster galaxies and determine
whether they define a tight color-magnitude relation over a wide range
in luminosity.  We then compare the colors and spectral diagnostics of
the different galaxy classes to identify any trends with, $e.g.$ age.

The paper is organized as follows: In \S2, we describe the HST imaging
and Keck spectroscopy, the completeness of our spectroscopic survey,
and define the spectral types.  We determine MS1054's mean redshift,
velocity dispersion, virial mass, and degree of substructure in \S3.
We analyse the color-magnitude diagram in \S4 and the composite
spectra in \S5.  We discuss how the red sequence forms in \S6 and
summarize our results in \S7.  Throughout the paper, we use
$\Omega_M=0.3,~\Omega_{\Lambda}=0.7$, and $H_0=70h$\kms~Mpc$^{-1}$.
At $z=0.83$, this corresponds to a projected scale of
$7.6$\hi~kpc~arcsec$^{-1}$.  Note that we use a different value for
$H_0$ than in our earlier papers on MS1054.

\section{Observations}

\subsection{HST/WFPC2 Imaging}\label{wfpc2}

In May 1998, a $\sim5'\times6'$ mosaic of the MS~1054--03 field was
taken with WFPC2 on the {\it Hubble Space Telescope}.  Six slightly
overlapping pointings were taken in both the F606W and F814W filters.
A full explanation of the WFPC2 image reduction is described in
\citet[hereafter vD00]{vandokkum:00}.

Because our spectroscopic survey was designed primarily from the WFPC2
imaging, we use the WFPC2 photometry to determine our spectroscopic
completeness (\S\ref{biases}).  We measure total magnitudes and galaxy
colors using SExtractor \citep{bertin:96} where galaxy colors were
measured within a $1''$ diameter apertures; note this is different
from vD00 where colors were measured within one effective radius.  At
$z=0.83$, $V_{606}$ and $I_{814}$ approximately correspond to
redshifted $U$ and $B$.  We stress that we only use the WFPC2
photometry in \S2.3.  For the remainder of the paper, we use the
magnitudes and colors measured from the ACS imaging.

\subsection{HST/ACS Imaging}\label{acs}

For the most of our analysis, we use the photometry and morphological
information derived from imaging taken with the HST {\it Advanced
Camera for Surveys} \citep{ford:98}.  A $2\times2$ overlapping mosaic
($5.8'\times5.8'$) was taken in F775W and F850LP in December 2002, and
the same pattern executed in F606W in January/February 2004.  A
detailed description of the ACS image reduction is presented in
\citet[hereafter B06]{blakeslee:06}.  To summarize, B06 fitted the 2-D
surface brightness distribution of each galaxy using a S\'ersic
profile; they then measured AB magnitudes and colors within the
effective radius.  We use their AB magnitudes and galaxy colors in our
analysis; we also adopt their nomenclature and refer to the AB
magnitudes in F775W as $i_{775}$, and the galaxy color corresponding
the difference between F606W and F775W as $(V-i_{775})$.  The
characteristic AB magnitude $m^{\ast}$ for the cluster population is
$i_{775}=22.3$ \citep{goto:05}.

All galaxies on the ACS mosaic brighter than $i_{775}=23.5$ were
visually classified by \citet[hereafter P05]{postman:05}.  Many of the
same galaxies also were classified earlier by vD00 using the WFPC2
imaging ($I_{814}<22$).  The WFPC2 imaging revealed a large number of
galaxy-galaxy mergers \citep{vandokkum:99} that have since been
spectroscopically confirmed by \citet{tran:05b}.  With the higher
resolution ACS imaging, P05 were able to morphologically classify the
galaxies making up these mergers, most of which are bulge-dominated
systems.  We direct the reader to P05 for a thorough discussion on the
morphological classifications and comparisons between the two surveys.
For our analysis, we use the Hubble types assigned by P05 but also
note the merging systems identified by vD00.  P05 assigned the
following morphological types: elliptical ($-5\leq T\leq-3$); S0
($-2\leq T\leq0$); and spiral+irregular ($1\leq T\leq10$).  The Hubble
types as well as ACS photometry for the cluster galaxies are listed in
Table~1.

\subsection{Transforming to Rest-frame Johnson Filters}\label{filter}

Throughout most of this paper, we use observed AB colors and
magnitudes from ACS; as stated in B06, this enables us to preserve the
model-independence of our results.  However, we do use rest-frame
colors and magnitudes for part of our analysis, and we follow B06 to
transform between the observed ACS filters and redshifted Johnson
filters.  In B06, the ACS/WFC photometry is transformed by computing
standard UBV rest-frame colors and observed ACS/WFC system colors for
the same models redshifted to $z=0.83$.  From B06 and \citet{jee:05},
we transform observed $V_{606}$, $i_{775}$, and $z_{850}$ to
rest-frame Johnson colors and magnitudes using:

\begin{equation}
B_z = i_{775}-0.16(i_{775}-z_{850})^2 - 0.22(i_{775}-z_{850}) + 0.73
\end{equation}

\begin{equation}
(U-B)_z = 0.85(V_{606}-i_{775}) - 1.01
\end{equation}

\begin{equation}
(U-V)_z = 0.95(V_{606}-z_{850})- 0.91
\end{equation}

\noindent Thus to transform the color scatter in, $e.g.$ observed
$(V_{606}-i_{775})$ (hereafter we use $V$ to refer to $V_{606}$) to
Johnson $(U-B)_z$, we simply multiply by 0.85.

\subsection{Keck/LRIS Spectroscopy}

\subsubsection{Observations \& Reduction}\label{lris}

Using the {\it Low Resolution Imaging Spectogram}
\citep[LRIS;][]{oke:95} on Keck, we carried out a spectroscopic survey
of the MS1054 field from February 1997 to June 2001; the spectra were
collected during six observing runs.  To select targets for the
multi-slit masks, two object catalogs were used.  The first catalog
was created from a 900 second Keck $I$ image ($1''$ seeing) centered
on the Brightest Cluster Galaxy (BCG).  Objects were detected with
FOCAS \cite{valdes:82} and $I$ magnitudes determined for a $3''$
diameter aperture.  Spectra taken in February 1997 and February 1998
were selected from the Keck $I$ catalog and are of objects with
$I<20.5$; the BCG has $I=17.5$ mag.  No morphological selection was
applied.

After obtaining our wide-field HST/WFPC2 imaging in May 1998, a second
object catalog was generated.  Objects were detected and best
magnitudes determined from the WFPC2 $I_{814}$ mosaic using SExtractor
\cite{bertin:96}; the BCG has $I_{814}=19.5$ mag.  Spectroscopic
targets for runs in January, February, and March 1999 as well as June
2001 were selected from the HST/WFPC2 object catalog.  As in the
earlier runs, target selection was based primarily on magnitude
($I_{814}\leq23.5$) and not color; in only one mask was morphology
used to select targets as we wished to obtain redshifts of close
galaxy pairs.

A total of 20 multi-slit masks (field of view $\sim7'\times7'$)
containing 711 targets were observed; many objects were observed in
multiple masks.  The multi-slit masks included both masks designed to
measure redshifts ($t_{\rm exp}\sim2000$ sec) and masks to measure
internal velocity dispersions ($t_{\rm exp}\geq10000$ sec) with
redshift fillers; many of the targets were observed in multiple masks.
The slits were $1''$ wide and, depending on the grating used, the
spectral resolution (FWHM) ranged from $9-13$\AA~with higher spectral
resolution ($5-6$\AA) used for the dispersion masks.

To reduce the spectra, we follow \citet{tran:99} and use IRAF
\footnote{IRAF is distributed by the National Optical Astronomy
Observatories, which are operated by the Association of Universities
for Research in Astronomy, Inc., under cooperative agreement with the
National Science Foundation.} routines with custom software provided
by L. Simard and D. Kelson.  IRAF was used to subtract the bias, remove
the hot pixels, and correct the y-distortion.  The spectra were then
cleaned of cosmic rays using software written by L. Simard: cosmic
rays were detected by determining a median value of a rectangular
region surrounding each pixel and using a rejection threshold of
$10\sigma_{CCD}$.  This approach circumvents spurious removal of
strong night sky lines.  

The spectra were flat-fielded, rectified, wavelength calibrated, and
sky-subtracted using Expector \citep{kelson:98}.  Because the
Tektronix CCD has severe fringing at $\lambda>7500$~\AA, multiple
flat-fields were taken throughout each night; the flat-fields
producing the lowest residuals were used.  The wavelength solution for
each slitlet was determined from the night sky lines; the list of
night sky lines from \cite{osterbrock:96} was used.  The spectra were
corrected for the telluric atmospheric A and B bands by using the
spectrum of a bright blue star included on all the masks; the blue
star also was used to roughly flux calibrate the data and remove fine
structure introduced by the grating.

The 1D spectra were extracted by summing the central five rows
(central $1.075''$) of each 2D spectrum.  The observed wavelength
coverage of the 711 targets depended on the grating and slitlet
position, but was typically $6000-9500$\AA.  For most cluster
galaxies, this includes the [OII]~$\lambda3727$ doublet and
4000\AA~break as well as the H$\delta$ and H$\gamma$ lines.

We refer the reader to \citet{tran:05a} for details on how redshifts
were measured using XCSAO \citep{kurtz:92}.  Typical redshift errors
from the cross-correlation routine were $\sim30$\kms.  However, these
quoted errors tend to underestimate the true errors because they do
not account for systematic uncertainties.  To better quantify the
errors, we compare the redshifts for 34 cluster galaxies that were
observed in multiple masks.  The median difference in the measured
redshifts for these galaxies is 0.0003 (50\kms).  

\begin{figure}
\plotone{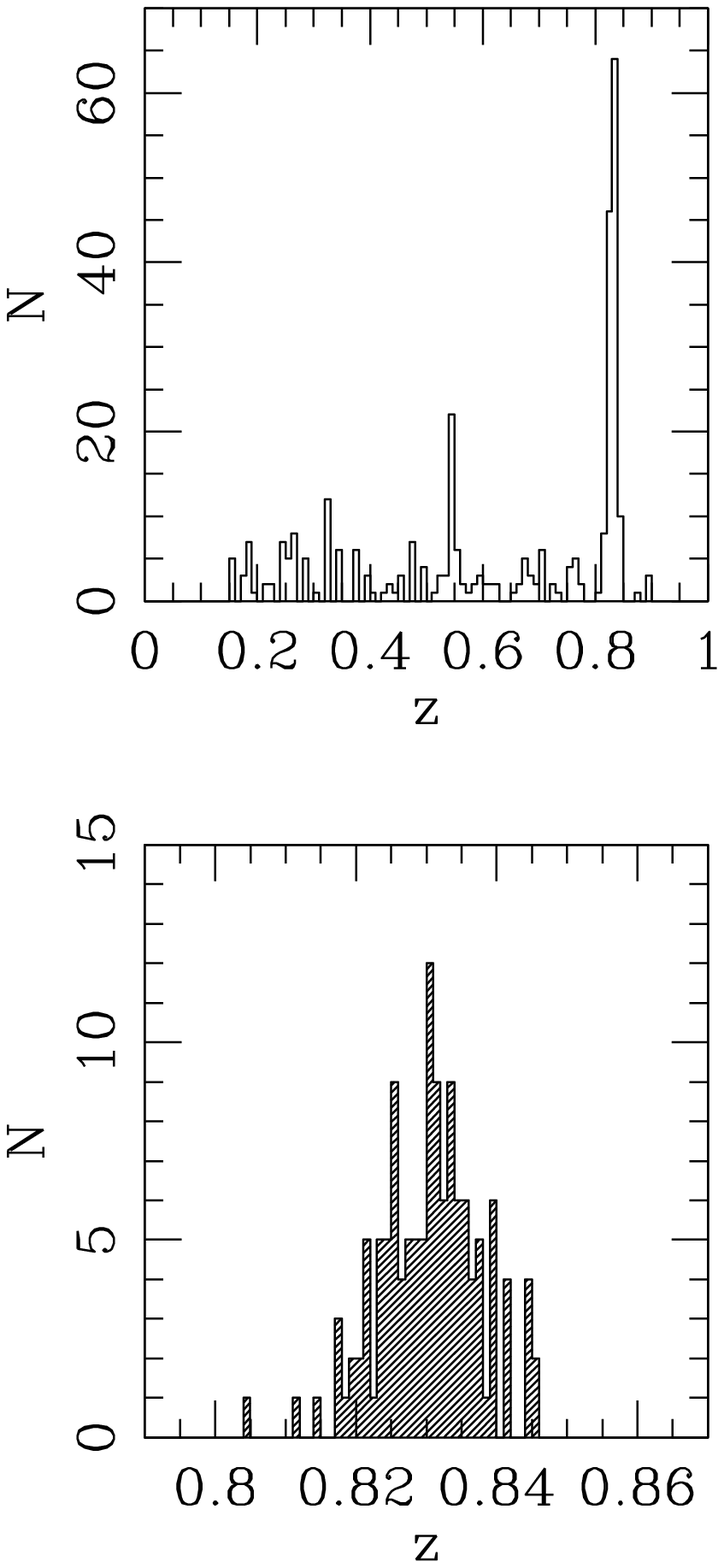}
\caption{From our catalog of 433 redshifts in the MS1054 field, we
 show the redshift distribution of the 307 galaxies at $0<z<1$ with
 quality flags of 3 (top), and an expanded view of the 129 galaxies
 ($Q_z=3$) that are in MS1054 (bottom).  Using these 129 members,
 MS1054's mean redshift and velocity dispersion are
 $z=0.8307\pm0.0004$ and $\sigma_z=1156\pm82$\kms, respectively.
\label{histz}}
\end{figure}

Including serendipitous objects, the final catalog has redshifts for
433 unique objects: 31 stars, 153 cluster galaxies, and 249 field
galaxies.  The redshift distribution for the galaxies at $0<z<1$ are
shown in Figure.~\ref{histz}.  All redshifts were given a quality flag
where $Q_z=3$ denotes multiple features (``definite''), $Q_z=2$
denotes redshifts with one strong feature and a weaker one
(``probable''), and $Q_z=1$ denotes only one feature (``estimate'').
In our analysis, we consider only members with a redshift quality flag
of $Q_z=3$; this reduces the total cluster sample to 129 members, of
which 121 fall on the HST/WFPC2 mosaic and 120 on the HST/ACS mosaic
(Table~1).  We also list the 24 galaxies that are likely to be MS1054
members but that have $Q_z=2$ or 1 in Table~1.  Redshifts for the 31
stars and 249 field galaxies will be published in Magee et al. (2007).

\subsubsection{Completeness \& Selection Effects}\label{biases}

In the following discusson on completeness and selection effects, we
consider only the objects that fall on the WFPC2 mosaic because most
of the spectroscopic survey was designed using the WFPC2 catalog (see
\S\ref{wfpc2}).  We consider the possibility of 1) magnitude bias due
the inherent difficulty of measuring redshifts for fainter objects and
2) color bias due to redshifts being more easily measured for galaxies
with emission lines.

\begin{figure}
\plotone{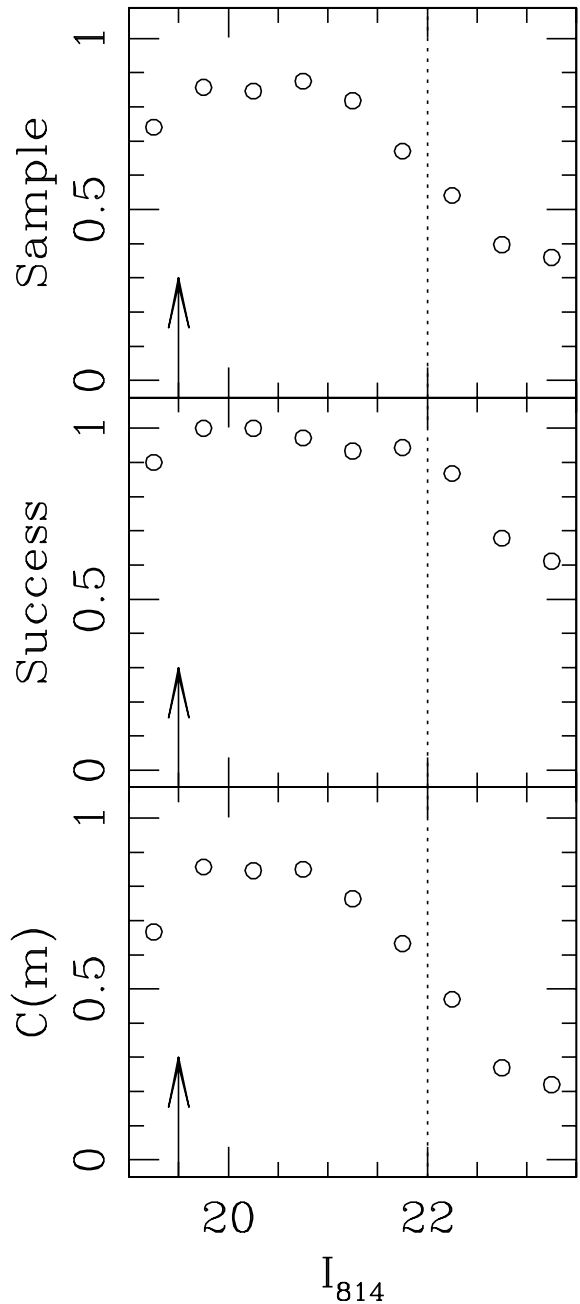}
\caption{{\it Top:} The sampling rate, defined as the number of
spectroscopic targets divided by the number of galaxies in the
photometric catalog, is shown as a function of magnitude (bin size
$\pm0.5$ mags); we consider only objects on the WFPC2 mosaic.  The
BCG's magnitude ($I_{814}=19.5$) is indicated by the arrow. {\it
Middle:} The number of acquired redshifts divided by the number of
targets: at $I_{814}=22$ (dotted line), the success rate remains
$\sim90$\%.  {\it Bottom:} The completeness, defined as the number of
redshifts divided by the number of galaxies in the photometric
catalog: incompleteness at the faint end is due to sparse sampling and
not the inability to measure redshifts of targeted galaxies.
\label{rates}}
\end{figure}

By comparing the number of galaxies in the WFPC2 photometric catalog
to the number of spectroscopic targets and acquired redshifts, we
investigate the completeness of our survey by following the analysis
described in \citet{yee:96}.  Figure~\ref{rates} shows the sampling
rate, success rate, and completeness of our sample as a function of
$I_{814}$.  The sampling rate, defined as the number of spectroscopic
targets divided by the number of galaxies in the WFPC2 catalog,
remains $>50$\% to $I_{814}=22$.  The success rate, defined as the
number of acquired redshifts divided by the number of targets, is even
higher at $\gtrsim90$\% to $I_{814}=22$.

The magnitude selection function $C(m)$, also known as the
completeness, is a measure of the magnitude bias and is shown in the
bottom of Fig.~\ref{rates}.  $C(m)$ is the product of the sampling and
success rates, and it is defined as the number of redshifts divided by
the number of galaxies in the photometric catalog.  Redshifts for 8 of
the 26 objects brighter than the BCG were not obtained because of a
geometric bias against objects at the edges of the WFPC2 mosaic.  As
we focus on cluster galaxies and these eight are clearly not members,
this is is not a concern.  Incompleteness at $I_{814}>22.0$ is due
mainly to sparse sampling rather than an inability to measure
redshifts at these magnitudes; this is a common characteristic of
redshift surveys \citep{yee:96,fisher:98}. For the cluster population,
we are $>75$\% complete at $I_{814}<=21.2$
\citep[$m^{\ast}$;][]{hoekstra:00}

\begin{figure}
\plotone{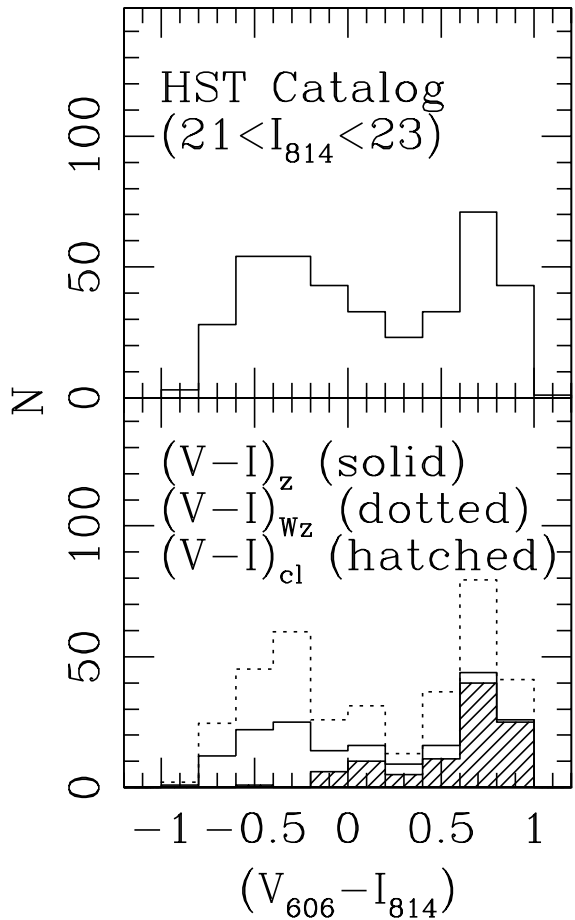}
\caption{The color distribution $(V_{606}-I_{814})$ of all galaxies in
the WFPC2 catalog with $21<I_{814}<23$; the bin size is $0.2$.  {\it
Bottom:} The $(V_{606}-I_{814})_z$ distribution for all galaxies on
the WFPC2 mosaic with redshifts (solid line) as well as cluster
members on the mosaic (hatched) in the same magnitude range.  We
include the weighted color distribution $(V_{606}-I_{814})_{Wz}$
(dotted) of the redshift sample where each galaxy is weighted by the
inverse of the magnitude selection function $C(m)$.  A K-S test finds
$(V_{606}-I_{814})_{Wz}$ is indistinguishable from
$(V_{606}-I_{814})$, $i.e.$ there is no measurable bias against faint
red (passive) galaxies in the spectroscopic sample after correcting
for sparse sampling.
\label{ri_hist}}
\end{figure}

To determine if our redshift sample is biased against faint, red
(passive) galaxies, we compare $(V_{606}-I_{814})$ of the galaxies in
the WFPC2 catalog to the subset with measured redshifts; because this
is more of a concern at fainter magnitudes, we focus on objects with
$21<I_{814}<23$.  Figure~\ref{ri_hist} shows the $(V_{606}-I_{814})$
distributions for galaxies in the WFPC2 catalog, the subset with
measured redshifts, and the cluster galaxies.  To remove the bias
introduced by sparse sampling at fainter magnitudes, we also determine
the weighted color distribution of the redshift sample
$(V_{606}-I_{814})_{Wz}$.  In this distribution, each galaxy is
weighted by the inverse of the magnitude selection function $C(m)$.
Applying the Kolmogorov-Smirnov test \citep{press:92} shows that
$(V_{606}-I_{814})_{Wz}$ is indistinguishable from $(V_{606}-I_{814})$
for galaxies in the WFPC catalog.  After correcting for sparse
sampling, there is no measurable bias against faint, passive galaxies
in our spectroscopic survey.

\subsubsection{Equivalent Widths \& Spectral Types}\label{spectypes}

We measure the equivalent widths (EW) of [OII]$\lambda3727$,
H$\delta$, and H$\gamma$ using the same spectral bandpasses used in
\citet{fisher:98} and \citet{tran:03b,tran:05a}.  Although H$\beta$ is
included in the wavelength range for most of the cluster galaxies, in
most cases it is compromised by strong sky lines.  The bandpasses are
based on \citet{worthey:97}, but the continuum sidebands have been
moved further away from the line center because these sidebands
overlap with the absorption lines for post-starburst galaxies.  In our
analysis, we adopt the convention that emission lines have negative EW
values and absorption lines have positive EW values.

\begin{figure}
\plotone{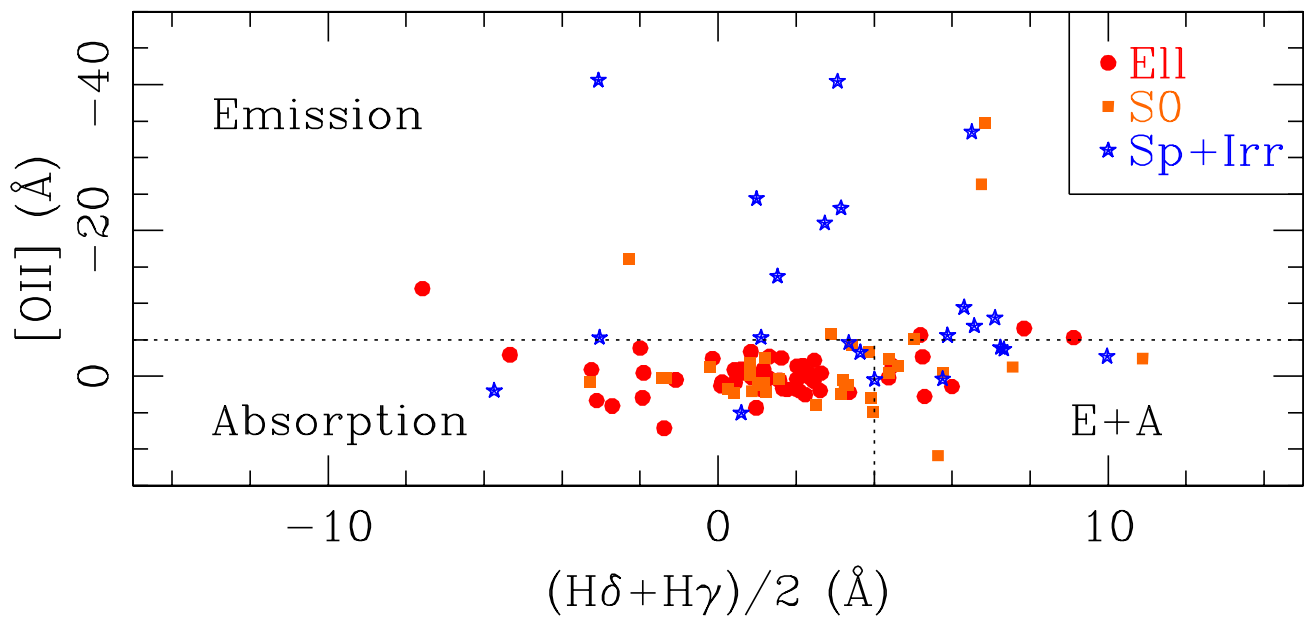}
\caption{The distribution of [OII]$\lambda3727$ versus Balmer line for
  115 cluster galaxies (five of the members with ACS imaging do not
  have measured [OII]).  The dotted lines denote the division between
  the three spectral types: emission-line, absorption-line, and
  post-starburst (E+A) galaxies.  The symbols denote the three
  morphological classes assigned by P05: elliptical ($-5\leq
  T\leq-3$), S0 ($-2\leq T\leq0$), and spiral+irregular ($T\geq1$).
\label{OII_balmer}}
\end{figure}

We separate the cluster galaxies into three spectral types: 1)
absorption-line members with no significant [OII] emission ($>-5$\AA)
and no strong Balmer absorption [(H$\delta+$H$\gamma)/2<4$~\AA]; 2)
emission-line members with strong [OII] emission ($\leq-5$\AA); and 3)
post-starburst (``E+A'') members that have very weak or no [OII]
emission ($>-5$~\AA) and strong Balmer absorption
[(H$\delta+$H$\gamma)/2\geq4$~\AA].  The measured equivalent widths
are listed by spectral type in Table~1.  Figure~\ref{OII_balmer} plots
[OII] versus Balmer values for 115 members (five of the 120 members
with ACS imaging do not have measured [OII]).

\section{Cluster Characteristics}

\subsection{Redshift, Velocity Dispersion, \& Virial Mass}

In our spectroscopic catalog of 433 objects in the MS1054 field, we
identify 153 galaxies with redshifts $0.80<z<0.86$; considering only
the cluster galaxies with redshift quality flags of $Q_z=3$ leaves 129
individual members.  The cluster galaxies define a strong peak in
redshift space (Fig.~\ref{histz}).

Using the biweight and jacknife methods \citep{beers:90}, we measure
the mean redshift and line of sight velocity dispersion of the 129
members to be $z=0.8307\pm0.0004$ and $\sigma_z=1156\pm82$\kms.
MS1054's velocity dispersion and X-ray temperature \citep[$8.6\pm1.1$
keV;][]{gioia:04} are consistent with no evolution in the
$\sigma_z-T_X$ relation to $z=0.83$.

To estimate MS1054's virial mass, we follow \citet{ramella:89} and
first determine the cluster's virial radius using:

\begin{equation}
R_V = \frac{\pi\bar{V}}{H_0} 
\sin \left\{ \frac{1}{2} \left[ \frac{N_m(N_m-1)}{2} \left( \sum_{i}
      \sum_{j>i} \theta_{ij}^{-1} \right)^{-1} \right] \right\}
\end{equation}

\noindent where $\bar{V}$ is the mean velocity of the cluster, $N_m$
is the number of members, and $\theta_{ij}$ the angular separation
between the $i^{th}$ and $j^{th}$ members.  With $R_{V}$, we can then
estimate the mass using

\begin{equation}
M_V = \frac{6\sigma_z^2 R_V}{G}
\end{equation}

\noindent where $\sigma_z$ is the line of sight velocity dispersion
and $G$ the gravitational constant.  

Using the 129 members, MS1054's virial radius is 1.8\hi~Mpc ($231''$)
and, using a bootstrap to resample the data-set 1000 times, we
estimate the associated error to be 0.1 Mpc.  The corresponding virial
mass is $3.3\times10^{15}M_{\odot}$ and, adding the errors due to
measuring the velocity dispersion and $R_{V}$ in quadrature, the total
error on the virial mass is $0.6\times10^{15}M_{\odot}$.  Our result
is consistent with both the weak-lensing and X-ray analyses by
\citet{jee:05}: assuming a single isothermal sphere, they measure a
virial radius from the ACS weak-lensing map of $1.5\pm0.1$\hi~Mpc and
from the $Chandra$ X-ray map of $1.7\pm0.2$\hi~Mpc.

\subsection{Substructure}\label{dstest}

\begin{figure}
\plotone{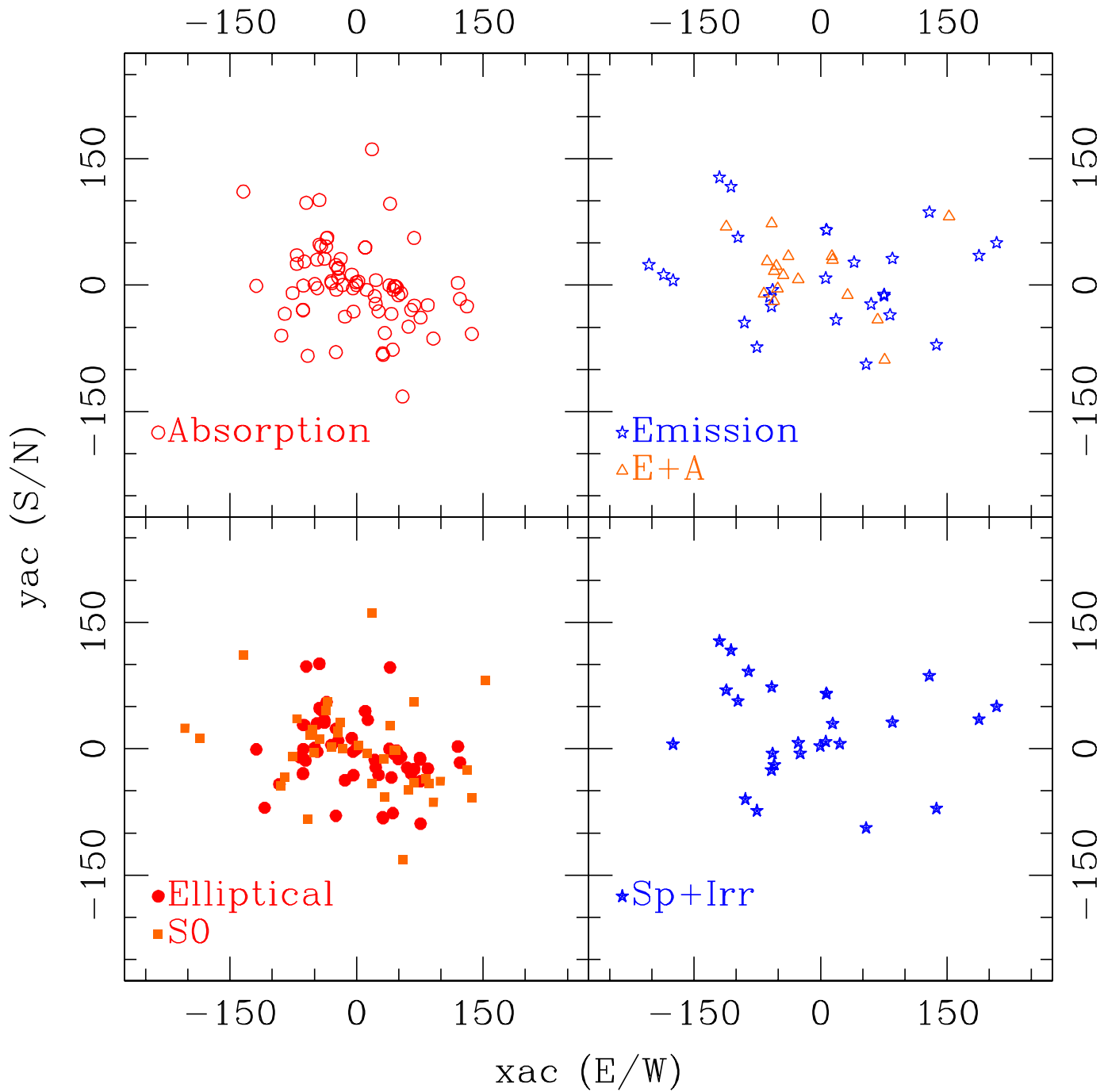}
\caption{Spatial distribution of the 120 galaxies in MS1054 with ACS
  imaging. The coordinates are relative to the BCG (H4520) and are in
  arcseconds; the $x$-axis corrresponds to right ascension ($-$east,
  $+$west) and the $y$-axis to declination ($-$south, $+$north).  The
  cluster galaxies are separated by spectral (top panels) and
  morphological class (bottom panels).  
\label{xydist}}
\end{figure}

\begin{figure}
\plotone{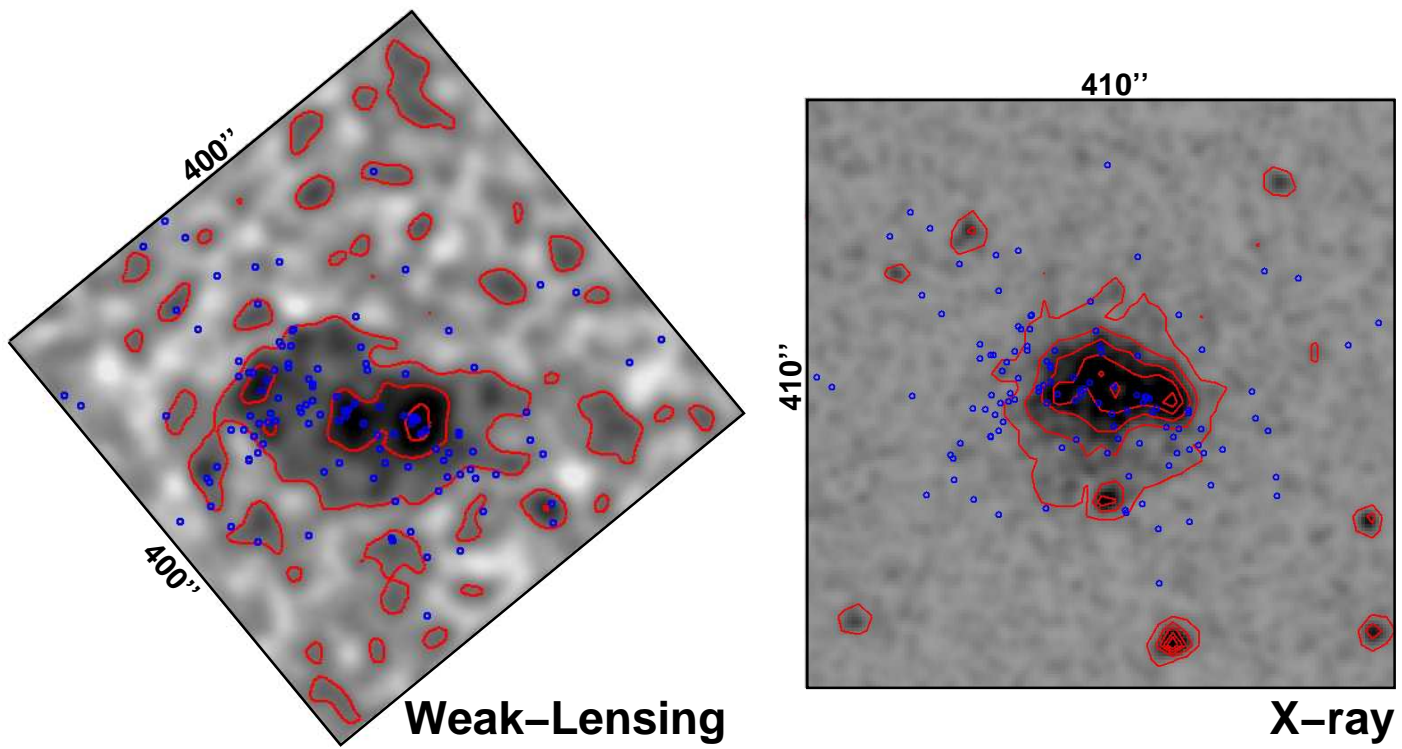}
\caption{{\it Left:} The ACS weak-lensing map of MS1054 from
  \citet{jee:05} is 400'' on a side where north is up and east to the
  left, as in Fig.~\ref{xydist}.  Mass contours corresponding to
  $\kappa=0.1, 0.2,$ \& 0.3 are shown in red, and the positions of the
  120 confirmed members that fall on the ACS imaging are shown as blue
  circles. {\it Right:} The smoothed XMM-$Newton$ map of MS1054 from
  \citet{gioia:04} is approximately 410'' on a side and has the same
  orientation.  The X-ray contours correspond to flux observed in the
  $0.5-10$~keV band, and the same cluster galaxies are shown.  For a
  color image of the MS1054 ACS mosaic, we refer the reader to
  \citet{blakeslee:06}.
\label{overlay}}
\end{figure}

Although MS~1054--03 falls on the $\sigma_z-T_X$ relation, it is not a
virialized structure: X-ray studies have shown that the cluster
contains two bright clumps, and weak-lensing analyses have revealed at
least three distinct mass peaks
\citep[$e.g.$][]{hoekstra:00,jeltema:01,gioia:04,jee:05}.  The spatial
distribution of the confirmed members is shown in Fig.~\ref{xydist}
where we have separated members by galaxy class.  We find that the
cluster galaxies trace the same elongated and clumpy structure that is
seen in both the X-ray and weak-lensing mass maps; in
Fig.~\ref{overlay}, we overlay the cluster galaxies onto the ACS
weak-lensing mass map from \citet{jee:05} and the XMM-$Newton$ X-ray
map from \citet{gioia:04}

To quantify the degree of substructure in MS1054, we use the
Dressler-Shectman test \citep{dressler:88}.  By using redshifts and
spatial positions, the D-S test quantifies how much the local mean
redshift and velocity dispersion (as defined by the ten nearest
neighbors to each galaxy) deviate from the cluster's global values.
Using biweight measures of redshift and velocity dispersion in the D-S
test, we find that the degree of substructure in MS1054 is significant
at the $>95$\% level.

\section{Color-Magnitude Relation}\label{cmr}

The color-magnitude diagram is a powerful tool for tracing the
relative (stellar) ages and the homogeneity of the different galaxy
populations \citep[$e.g.$][]{bower:92}.  In our analysis, we consider
only the 120 members that lie on the ACS mosaic with redshift quality
flags of 3 (see \S\ref{lris}).  This ensures that we include only the
cluster galaxies with measured colors, magnitudes, and Hubble types
from the ACS imaging.  We group the members into the three spectral
types defined in \S\ref{spectypes} as well as the three morphological
types classified by P05: elliptical ($-5\leq T\leq-3$), S0 ($-2\leq
T\leq0$), and spiral+irregular ($T\geq1$).  The spatial distributions
of the different galaxy classes are shown in Fig.~\ref{xydist}.  Of
these 120 members, 115 are also assigned spectral types.

\subsection{Fitting the Color-Magnitude Relation}

\begin{figure}
\plotone{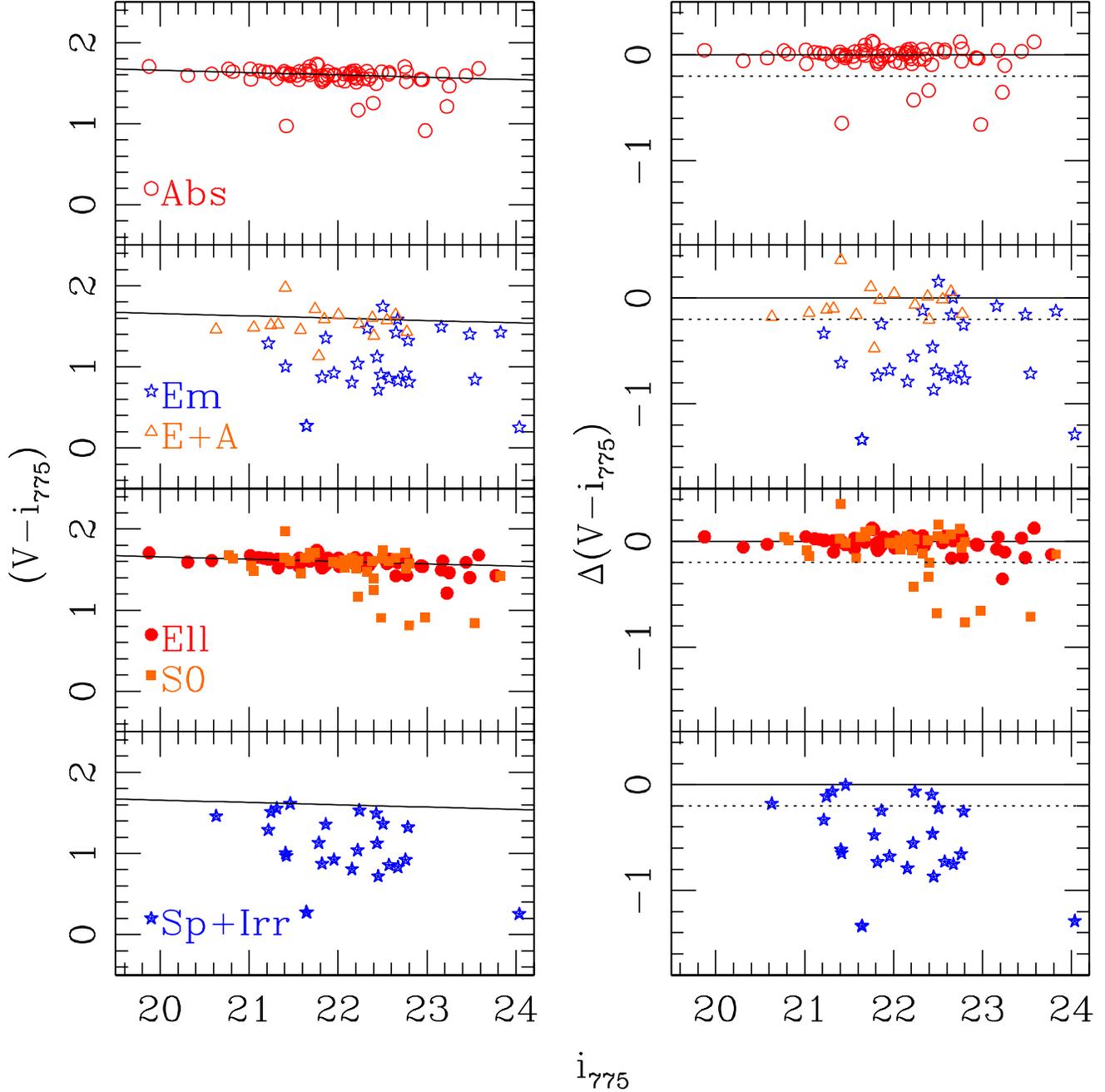}
\caption{{\it Left panels:} Color-magnitude (CM) diagrams for the
  cluster galaxies that fall on the ACS mosaic
  \citep[$i_{775}^{\ast}\sim22.3$;][]{goto:05}.  In the top two
  panels, the members are separated into the three spectral types:
  absorption-line (open circles), emission-line (open stars), and
  post-starburst (E+A; open triangles) galaxies.  In the bottom two
  panels, the members are separated into the three morphological types
  as classified by \citet{postman:05}: ellipticals (filled circles),
  S0s (filled squares), and spiral+irregulars (filled stars).  The
  solid line in all four panels is the CM relation determined by
  fitting the 67 red absorption-line members.  {\it Right panels:} For
  the same galaxy classes, the difference between their measured color
  and the color predicted from the CM relation; the dotted line
  denotes $\Delta(V-i_{775})=-0.2$, our division between red and blue
  members.  The absorption-line members define a strikingly narrow
  sequence along the fitted CM relation: 67/72 have
  $\Delta(V-i_{775})>-0.1$.
\label{cmd.vi}}
\end{figure}

In Fig.~\ref{cmd.vi} (left panels), we show the color-magnitude (CM)
diagram where the 115 (120) members are separated into the different
spectral (morphological) types.  Here we use the AB $i_{775}$
magnitudes and $(V-i_{775})$ colors measured by B06.  The colors are
measured within a circular effective radius, and the measurement
errors on the colors are $\delta(V-i_{775})\sim0.02$ for members
brighter than $i_{775}=22.5$ and $\delta(V-i_{775})\sim0.03$ for
fainter members.

To fit the CM relation and determine the scatter in color, we use the
absorption-line members.  From Fig.~\ref{cmd.vi} (left panels), we see
that these members define a narrow sequence over $\sim3.5$ magnitudes:
their deviation from the CM relation (quantified as
$\Delta(V-i_{775})$) measured by B06 is less than 0.2.  To determine
the CM relation defined by the red absorption-line members, we exclude
the five galaxies that clearly deviate from the CM relation and use
the remaining 67 absorption-line members to determine that:

\begin{equation}
(V-i_{775}) = 1.602 - 0.024(i_{775}-22)
\end{equation}

\noindent Compared to B06, who fit the CM relation using only the
ellipticals, we find a slightly flatter slope; however, this
difference is within the errors (B06 quote an error of 0.014, and we
estimate a comparable error).  If we use only the 62 absorption-line
members with $|\Delta(V-i_{775})|<0.1$, the slope is $-0.032$.

In Fig.~\ref{cmd.vi} (right panels), we show the difference between
the measured color and the color predicted from our fitted CM relation
for the different galaxy classes.  We consider
$\Delta(V-i_{775})=-0.2$ to be the division between red and blue
galaxies.  The measured offset and scatter in color for the different
galaxy classes are listed in Table~\ref{colortab}; the offset and
scatter were determined using the biweight estimate and scale,
respectively \citep{beers:90}.  Following B06, we also determine the
intrinsic scatter in color ($\sigma_{int}$) by correcting for
measurement errors.  Included in Table~\ref{colortab} are the mean
$(V-i_{775})$ and $(i_{775}-z_{850})$ colors, where both are also
determined using the biweight estimate.  The uncertainties in the
offset, scatter, and colors were determined by bootstrap resampling
(biweight estimate of 5000 realizations).  Changing the slope, $e.g.$
from $-0.024$ to $-0.032$, has neglible impact on the measured color
offsets and scatters; the differences are within the quoted errors.

\begin{figure}
\plotone{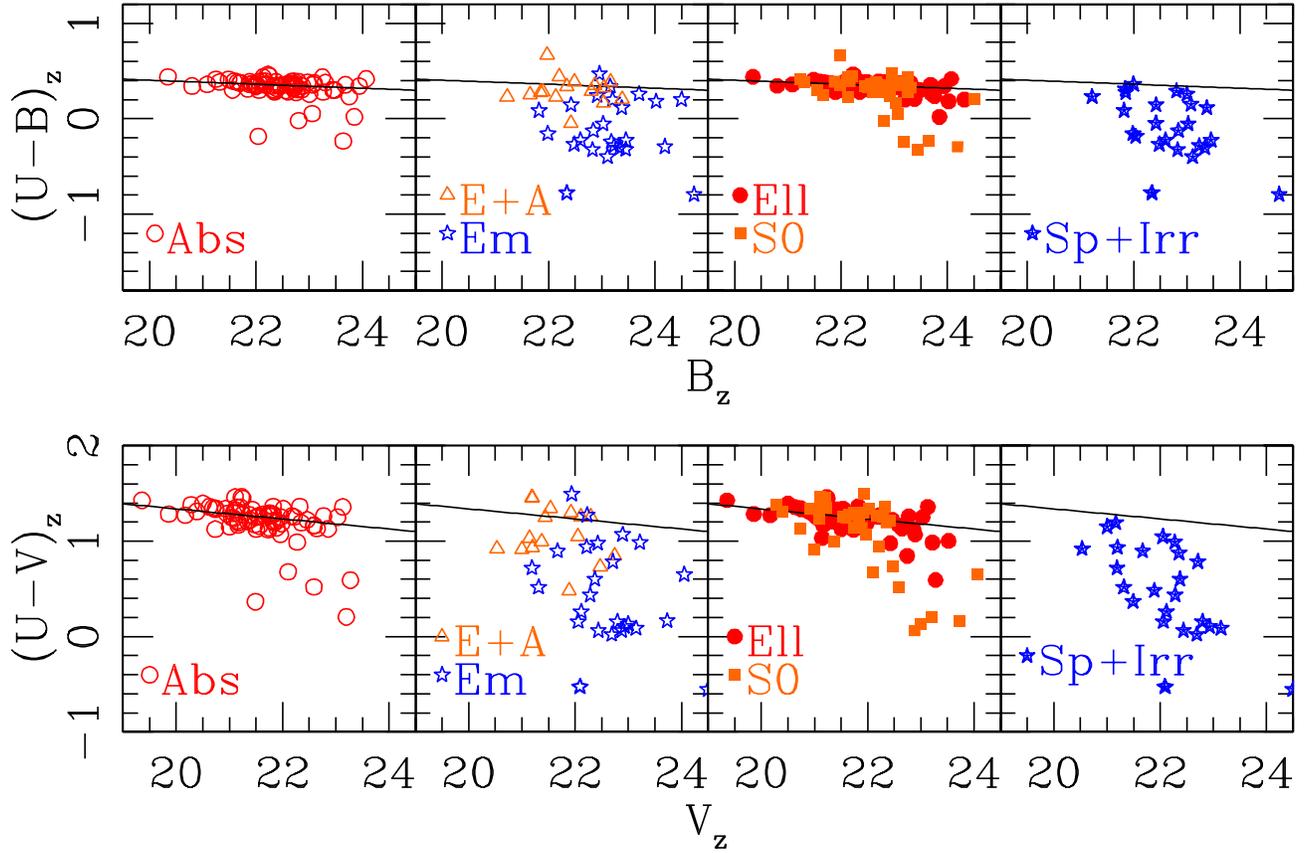}
\caption{{\it Top panels:} CM diagrams for the same galaxy classes as
  in Fig.~\ref{cmd.vi} but in rest-frame colors corresponding to
  $(U-B)_z$ (top panels) and $(U-V)_z$ (bottom panels).  The spectral
  types (absorption, E+A, emission) are always open symbols while the
  morphological types (E, S0, Sp+Irr) are always solid symbols.
  Again, the absorption-line members have the lowest scatter in color
  of all the galaxy types.
\label{cmd.rest}}
\end{figure}

At $z=0.83$, the observed $(V-i_{775})$ and $(V-z_{850})$ colors
correspond very closely to Johnson $(U-B)_z$ and $(U-V)_z$.  For
comparison to earlier work on Coma \citep{bower:92,terlevich:01} and
MS1054 (vD00), we include CM diagrams for these rest-frame colors
(Fig.~\ref{cmd.rest}).  Again we fit the CM relation using the 67 red
absorption-line members:

\begin{equation}
(U-B)_z = 0.364 - 0.021(B_z-22)
\end{equation}

\begin{equation}
(U-V)_z = 1.232 - 0.053(V_z-22)
\end{equation}

\noindent The CM relation in $(U-B)_z$ is consistent with that
determined by vD00 for MS1054 using WFPC2 photometry of 30 E+S0
members.  As noted in \S\ref{filter}, the color scatter in observed
$(V-i_{775})$ is easily transformed to $(U-B)_z$ by multiplying by
0.85.

\subsection{Absorption-Line Members:  A Uniformly Tight Red Sequence}

In our spectroscopic sample, $63\pm7$\% of the members are absorption
line galaxies.  The absorption-line members define a strikingly narrow
sequence along the CM relation: virtually all of these galaxies
(67/72) have $\Delta(V-i_{775})>-0.2$; the few that are blue are
consistent with the associated error on the measured absorption-line
fraction.

The intrinsic scatter in color for the absorption-line members is
extremely small; it is comparable to that of the morphologically
classified ellipticals ($\sigma_{Vi}=0.048\pm0.009$
vs. 0.055$\pm0.008$; Table~\ref{colortab}).  We note that most
galaxy cluster surveys at $z>0.5$ do not have the high resolution
space-based imaging needed to separate ellipticals from S0s.  If we
consider instead the ellipticals and S0s together as a single class,
their intrinsic color scatter ($\sigma_{Vi}=0.072\pm0.010$) is
measurably larger than that of the absorption-line members.

For the red absorption-line members, the scatter in color remains
small over the observed magnitude range: it is essentially the same
for both bright and faint members (Table~\ref{colortab}).  In
contrast, the scatter in color for the faint red ellipticals is
$\sim50$\% larger than for the bright red ellipticals.  In this
comparison, we have excluded any absorption or elliptical members with
$\Delta(V-i_{775})<-0.2$ to ensure that the color scatter is not
inflated by blue outliers

The tightness of the CM relation for the absorption-line members
demonstrates that their stellar populations are uniformly old; the
intrinsic scatter in $(V-i_{775})$ corresponds to $\sigma_{UBz}$ of
only $0.041$.  Comparing this small color scatter to the mixed star
formation models from \citet{vandokkum:98a} indicates that all star
formation ceased by $z\sim1.2$, and that most of these members
($>75$\%) formed by $z\sim2$.  This is consistent with the {\it mean}
stellar ages we estimate from their colors: comparing to the single
stellar burst models (solar metallicity) in B06, their mean
$(V-i_{775})$ and $(i_{775}-z_{850})$ colors correspond to mean
stellar ages of $\sim3$ Gyr.

In MS1054, we see that spectroscopy with even a generous color cut of
$\Delta(V-i_{775})>-0.2$ is extremely effective at identifying a
homogeneously old population that follows the fitted CM relation over
a factor of $\sim25$ in luminosity.  Our survey shows that selecting
by spectral type can be better at isolating a uniformly old galaxy
population than with only either color or morphological cuts,
especially over a wide range in luminosity at $z>0.8$.

\subsection{Post-Starburst (E+A) Members}

Post-starburst galaxies, often referred to as E+A or k+a galaxies, are
galaxies where star formation has been truncated.  E+As make up
$5-20$\% of the galaxies in $z>0.3$ clusters
\citep{dressler:83,couch:87,dressler:99,tran:03b}, but whether the E+A
phase plays an important role in the evolution of early-type members
remains a point of contention \citep{balogh:99,dressler:04}.  This is
primarily because E+As can only be identified spectroscopically and,
being a short-lived phenomenon \citep[$\lesssim1.5$ Gyr;][]{couch:87},
they are relatively rare.

In MS1054, we find that the E+As make up $15\pm4$\% of the cluster
population; this is consistent with the fraction published in
\citet{tran:03b}.  An E+A spectrum can be produced via several star
formation histories: while most commonly associated with a combination
of a starburst that ended within the last $\sim1.5$ Gyr and an older
stellar population
\citep{dressler:83,couch:87,barger:96,poggianti:99}, E+As can also
result from continuous star formation that was truncated
\citep{newberry:90}.  This combined with variations in the stellar
mass produced in the most recent star formation episode leads to E+As
that can span a range in {\it mean} stellar age, and thus the E+As
should have a larger scatter in color compared to the absorption-line
members.

The E+As in MS1054 have an intrinsic color scatter that is three times
larger than that of the absorption-line population:
$\sigma_{Vi}=0.152\pm0.052$ vs. $0.048\pm0.009$
(Table~\ref{colortab}).  However, the relative age differences and
corresponding scatter in color will diminish as the E+As evolve.
Assuming no further star formation, the E+As observed at $z=0.83$
($\sigma_{UBz}=0.129$) will lie on the CM relation and be
indistinguishable from the absorption-line members by the present
epoch \citep[$e.g.$][]{bower:98}.  Thus the current generation of E+As
will increase the absorption-line fraction by a substantial amount
($\sim15$\%).

One notable E+A is the unusually red S0 member H3910.  As discussed in
B06, H3910's colors suggest that there was starburst $\sim0.5$ Gyr ago
and that the galaxy is currently heavily obscurred by dust.  This is
consistent with the post-starburst nature of H3910's spectrum.  It is
also possible that the dust in H3910 is obscurring any current star
formation, as is observed in a handful of post-starburst galaxies in
a cluster at $z=0.41$ \citep{smail:99}.

For the remainder of the paper, we use the terms E+A and
post-starburst interchangeably.  However, we recognize that there is
some ambiguity given that 1) truncated star formation can also produce
an E+A and 2) even if there was a burst, there is no guarantee that
there will not be another one in the future.  A more generic term
would be post-starforming galaxy, but we use the adopted terminology
to emphasize the rapid truncation in star formation that is required,
regardless of the star formation history leading up to it.

\subsection{Emission-Line Members}

In MS1054, the emission-line members make up $23\pm4$\% of the cluster
population and they tend to have bluer colors than both the
absorption-line and E+A members.  However, not all of the
emission-line members are blue: 7/26 have $\Delta(V-i_{775})>-0.2$.
In addition, all of these red emission-line members are fainter than
$i_{775}=22$ and have early-type morphologies.  We note that a similar
population of emission-line, red, early-type members also has been
found in Cl0152 \citep[$z=0.837;$][]{homeier:05}.  Color cuts, even
when combined with morphological classifications, cannot separate
emission from absorption-line members, especially at fainter
luminosities.

\begin{figure}
\plotone{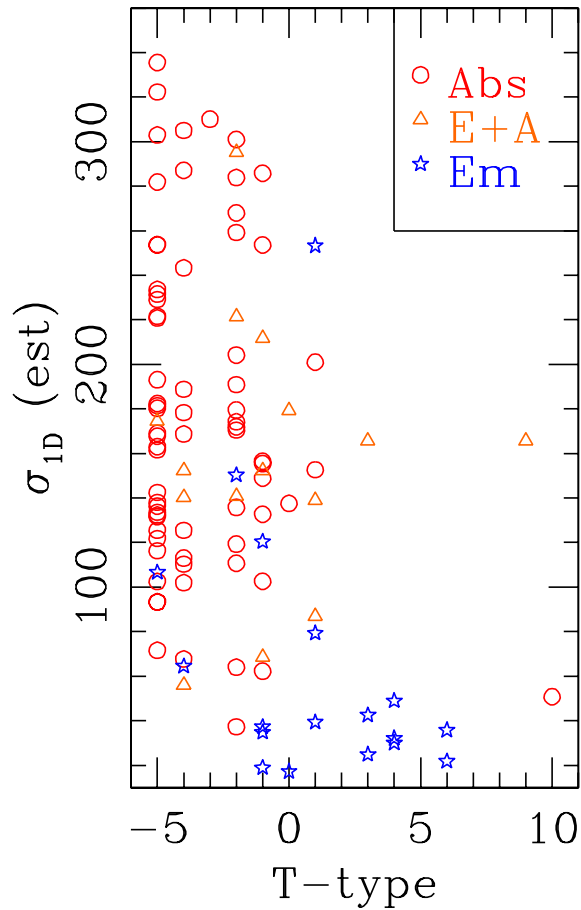}
\caption{The morphological types assigned by P05 versus internal
  velocity dispersions: the E+S0 members have $T\leq0$ while
  Spiral+Irregulars have $T\geq1$. Internal velocity dispersions have
  been measured for 27 of the members \citep[errors
  $<50$\kms;][]{wuyts:04}, and $\sigma_{1D}$ estimated for the
  remainder \citep[see][]{tran:03b}.  The symbols correspond to the
  spectral classifications.  Most of the emission-line members are low
  mass ($\sigma_{1D}<100$\kms) systems, and more than half are Sp+Irr.
  Assuming the emission-line members evolve onto the red sequence at
  $z<0.8$, they can only be low-mass, low-luminosity members.
\label{Ttype_nsigma}}
\end{figure}

The emission-line members have the largest color offset and scatter of
all the galaxy classes; their color scatter is even larger than that
of the morphologically classified Sp+Irr members ($0.407\pm0.064$
vs. $0.273\pm0.065$).  Comparing their mean colors
(Table~\ref{colortab}) to the solar metallicity models from B06
indicates that a typical emission-line member is $\sim1.5-2$ Gyr
younger than the absorption-line members and has had some combination
of continuous or bursting star formation.  Most of these galaxies
($>75$\%) will fade into faint ($L<L^{\ast}$) members.  This is
supported by their internal velocity dispersions ($\propto$mass):
$\sim80$\% of the emission-line members have estimated internal
velocity dispersions\footnote{Following \citet{tran:03a}, we use
measured internal velocity dispersions ($\sigma_{1D}$) for 27 of the
members \citep[errors $<50$\kms;][]{wuyts:04} and estimated
$\sigma_{1D}$ for the remainder.} of $<100$\kms~
(Fig.~\ref{Ttype_nsigma}).  Assuming they evolve into low-mass
early-type members, their younger mean stellar ages relative to the
more massive early-types are consistent with trends observed in lower
redshift clusters, $e.g.$ Coma and Abell 2218
\citep{poggianti:01b,smail:01}.

\subsection{The Morphological Types:  Elliptical, S0, \&
  Spiral+Irregular Members} 

Using the morphological types assigned by P05, we find that the
ellipticals in MS1054 define a tight red sequence that is comparable
to that of the absorption-line members (Fig.~\ref{cmd.vi};
Table~\ref{colortab}).  In comparison, the intrinsic color scatter for
the S0s is twice that of both the ellipticals and the absorption-line
members (Table~\ref{colortab}); this is true even if we compare only
the brightest ($i_{775}\leq22$) S0s to the brightest ellipticals and
absorption-line members.  However, the increased color scatter for the
S0 sample is primarily due to a small fraction of blue S0s; most of
the S0s (28/39; 72\%) are as red as the ellipticals and
absorption-line members.  If we consider only members with
$\Delta(V-i_{775})>-0.1$, the intrinsic color scatter for the
absorption-line, elliptical, and S0 samples are comparable:
$0.044\pm0.011$, $0.040\pm0.011$, and $0.047\pm0.018$, respectively.

The differences in the color scatters indicate that, on average, the
ellipticals are more uniform in age than the S0s.  Dust can also play
a role by making some of the S0s redder (as in the case of H3910; see
\S4.3); dust would decrease the mean color offset for the S0s, but the
S0s would still retain a larger color scatter.  Unlike the
ellipticals, a number of the S0s are blue (7/39; 18\%) and/or are
post-starburst systems (8/39; 21\%); this suggests that the S0s and
ellipticals are not drawn from the same parent population.  We return
to this issue in \S\ref{composite}.

Note that all seven of the blue S0s are fainter than $i_{775}=22$;
assuming these S0s will evolve onto the red sequence at $z<0.8$, they
will remain $L<L^{\ast}$ members.  The resulting younger mean ages for
the faint S0s is consistent with the difference in mean age between
the bright ellipticals and faint S0s in Coma \citep{poggianti:01b}.

As reported in vD00, the Spiral+Irregular members have the largest
color offset and scatter of the morphological classes
(Table~\ref{colortab}).  However, several of the bright Sp+Irr are red
and considered to be absorption-line or post-starburst systems (see
Fig.~\ref{cmd.vi}).  Again we see that the combination of color and
morphology does not consistently correspond to recent star formation:
many of the Sp+Irr members are red and/or absorption-line systems, and
vice versa for the S0 members.

\subsection{Color Scatter in Rest-frame Johnson Filters}

In our analysis, we prefer to use observed ACS colors to preserve the
model-independence of our results.  However, we include rest-frame
Johnson colors so that we can compare to the color scatter measured
in, $e.g.$ Coma.  In Table~\ref{uvztab}, we list the mean color offset
and color scatter in $(U-V)_z$ for the galaxy classes.  As described
earlier, the CM relation is fitted to the 67 red absorption-line
members.

The differences in the $(U-V)_z$ offset and scatter between the galaxy
classes are consistent with our earlier results using observed
$(V-i_{775})$: the absorption-line members define the tightest red
sequence with an intrinsic $\sigma_{UVz}=0.083\pm0.012$.  The
elliptical-only sample also has a small $(U-V)_z$ scatter of
$0.087\pm0.015$.  In comparison, the S0-only sample has a color
scatter that is nearly twice as large ($\sigma_{UVz}=0.187\pm0.063$);
this increases the $(U-V)_z$ scatter for the E+S0 population as a
whole to $0.110\pm0.020$.  However, we again note that the larger S0
scatter is mainly due to a number of S0s ($\sim20$\%) that are blue,
and that most of the S0s are as red (old) as the elliptical and
absorption-line members (see Fig.~\ref{cmd.rest}).

The absorption-line members in MS1054 have an intrinsic $(U-V)_z$
scatter that is more than twice as large as that of the Coma
ellipticals \citep[$\sigma_{UV}=0.036\pm0.020$;][]{terlevich:01}.  If
the current generation of absorption-line members in MS1054 continue
to evolve passively, their $(U-V)_z$ scatter will decrease
considerably over the next $\sim7$~Gyr, the look-back time to
$z=0.83$.  Comparing to the mixed star formation models from
\citet{vandokkum:98a}, we see that in models where all activity ceases
by $z\sim1.2$ the color scatter decreases by approximately a factor of
three between $z\sim0.8$ and $z\sim0$.  Thus by $z\sim0$, the
absorption-line members in MS1054 will have as small a color scatter
as the Coma ellipticals.  This decrease in the $(U-V)$ scatter is also
consistent with the mixed star formation histories for Coma modeled by
\citet{bower:98}.  The absorption-line galaxies in MS1054 will evolve
into the oldest members of local clusters.

\section{Composite Cluster Spectra}\label{composite}

\begin{figure}
\plotone{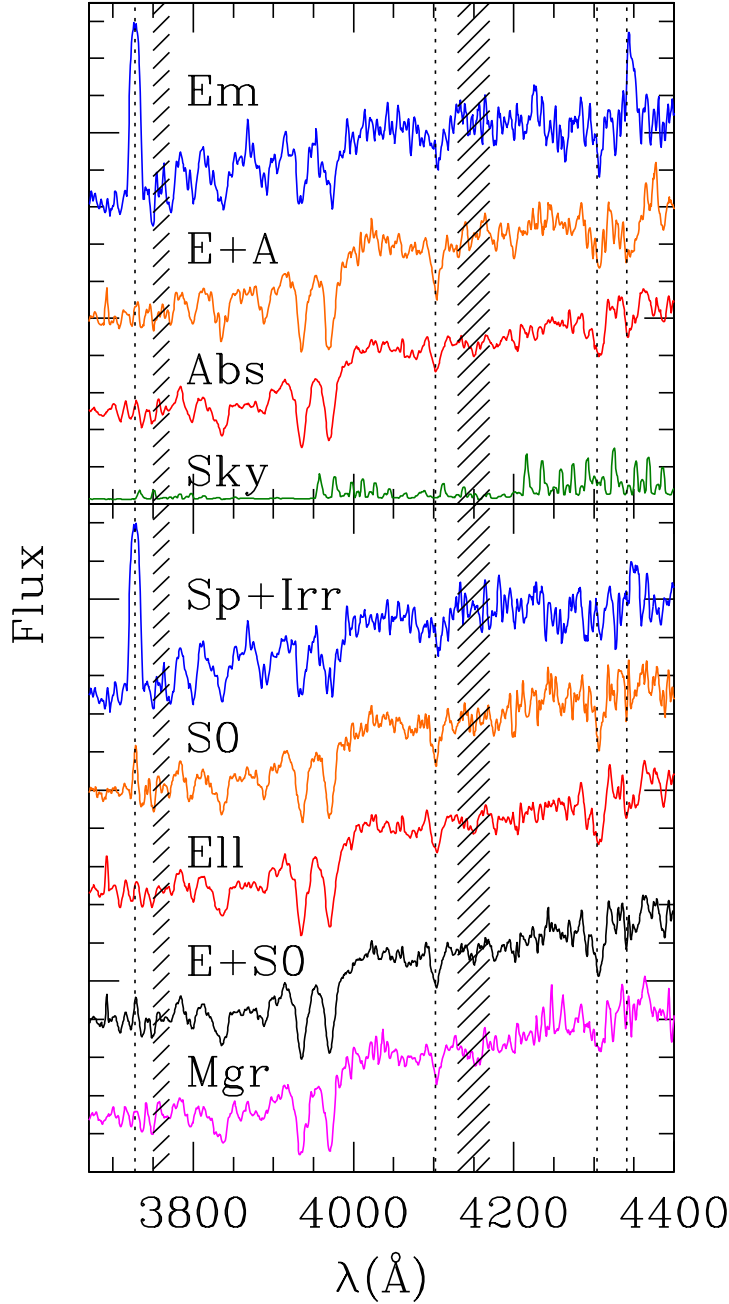}
\caption{{\it Top:} Smoothed composite spectra for the three different
  spectral types plotted in the rest-frame on an arbitrary flux scale;
  the dotted vertical lines denote [OII]$\lambda3727$, H$\delta$,
  G-band, and H$\gamma$.  We include a spectrum of the night sky-lines
  and indicate the noisier regions corresponding to corrections for
  telluric absorption (hatched bars); the observed wavelengths of the
  sky features have been divided by 1.83.  H$\delta$ absorption is
  evident in all three composite spectra but is weakest in the
  absorption-line spectrum (Table~\ref{indices}). {\it Bottom:}
  Smoothed composite spectra for the different morphological classes;
  we show a combined early-type spectrum (E+S0) as well as spectra for
  the individual types.  We also include the composite spectrum for
  the 17 members that are in red, merging systems from
  \citet{tran:05a}.  H$\delta$ absorption is evident in all of the
  composite spectra, even when only considering the ellipticals
  (Table~\ref{indices}).
\label{coadd}}
\end{figure}

To investigate the global properties of the cluster population, we
combine the spectra of the 120 members by galaxy type.  We first
normalize each 1D spectrum to unity over rest-frame $4200-4600$\AA~and
then take an average because we are interested in a representative
spectrum for each galaxy type.  We then remove spuriously high pixel
values due to, $e.g.$ contamination by sky lines, by smoothing the
average spectrum using a large window, subtracting this smoothed
spectrum, and removing pixel values that are 30\% larger.  The
composite spectra for the spectral and morphological classes are shown
in Fig.~\ref{coadd}.

For each composite spectrum, we measure the same spectral indices
defined in \S\ref{spectypes}.  We determine errors by bootstrapping
1000 realizations for each galaxy class using the base set, and then
measuring the same indices in the generated composite spectra; the
error is the biweight scale of the resulting distribution divided by
$\sqrt{N_g}$.

As part of our analysis, we also measure H$\delta_A$ and $D_N(4000)$
using the bandpasses defined in \citet[hereafter K03]{kauffmann:03}.
H$\delta_A$ and $D_N(4000)$ were developed by K03 as a method to
detect starbursts that occurred within the past $1-2$ Gyr and thus
better quantify the star formation histories of the local galaxies in
the Sloan survey.  With our composite spectra, we can robustly measure
both of these spectral diagnostics and compare them directly to K03's
library of different star formation histories generated using
\citet[hereafter BC03]{bruzual:03} models.  We list the measured
spectral indices in Table~\ref{indices}.

\subsection{The Spectral Types:  Absorption-line, Post-Starburst, \&
Emission-line}

In Fig.~\ref{coadd} (top), we show the composite spectra for the
absorption-line, post-starburst (E+A), and emission-line members.  The
composite spectra have the features that are typically associated with
the different types: the absorption-line spectrum has the strong
G-band absorption that results from a stellar population dominated by
F and G type stars \citep[$e.g.$ see][]{gunn:83}; the E+A spectrum is
similar to the absorption-line one, but with stronger H$\delta$ and
H$\gamma$ absorption due to more recent star formation ($<2$ Gyr); and
the emission-line spectrum shows [OII] emission that presumably is
associated with current star formation.

Using the Sloan survey, \citet{dressler:04} showed that the strong
H$\delta$ absorption coupled with no [OII] emission observed in
distant clusters cannot be reproduced by a mix of local star formation
histories, $i.e$ by star formation that is quenched over long ($>1$
Gyr) timescales.  Rather, more intense star formation such as bursts
are required.  Given their strong Balmer absorption, we expect such a
starburst phase for the E+As.  The emission-line members also show
strong Balmer absorption indicative of a recent spate of star
formation in addition to the current level of activity.  Even the
absorption-line members have weak H$\delta_A$ absorption
($\sim0.5$\AA), a value consistent with a mean stellar age of
$\sim2.8$ Gyr (KC03).

\subsection{The Morphological Types:  Elliptical, S0, \&
  Spiral-Irregular}\label{es0_spec}

In Fig.~\ref{coadd} (bottom), we show composite spectra for the
combined early-type sample as well as for the elliptical, S0, and
Spiral+Irregular members.  The combined early-type sample shows
H$\delta$ absorption that is measurably stronger than in the
absorption-line spectrum (H$\delta=2.5$ vs. 1.7; Table~\ref{indices}).
Assuming that the stronger H$\delta$ absorption is due to more recent
star formation, the morphologically classified E+S0 members are not as
homogeneously old as the absorption-line population.

Because the ACS imaging allows us to separate the ellipticals and S0s,
we can compare their individual composite spectra to search for any
differences in their mean stellar populations, $i.e.$ any trends
introduced by the blue S0s (see \S\ref{cmr}).  Note that K-S tests
confirm the ellipticals and S0s have indistinguishable $i_{775}$ and
$\sigma_{1D}$ distributions, $i.e.$ any differences between the two
composite spectra are unlikely to be due to other factors such as the
mass-metallicity relation.

The H$\delta_A$ absorption in the S0s is nearly $2$\AA~stronger and
their $D_N(4000)$ index lower than in the ellipticals.  Surprisingly,
the S0 spectrum also shows weak [OII] emission, a sign of ongoing star
formation.  These differences are consistent with the larger scatter
in color exhibited by the S0s compared to the ellipticals
(\S\ref{cmr}; Table~\ref{colortab}), and are due mostly to the S0s
that are blue (7/39; 18\%).  Comparing H$\delta_A$ and $D_N(4000)$ to
the single burst model (solar metallicity; KC03) shows an age
difference of $\sim0.5-1$ Gyr.

Is there still a difference between the ellipticals and S0s if we
exclude the seven faint, blue S0s?  To test this, we generate
composite spectra for the bright, red ellipticals and S0s
($i_{775}\leq22$; $\Delta(V-i_{775})\geq-0.2$; Table~\ref{indices}).
The mean luminosities of both sets are identical ($i_{775}=21.4$),
thus we can be sure that any differences in the spectral indices are
not due to a luminosity effect.  

A surprisingly high fraction (30\%; 4/13) of the bright, red S0s are
post-starburst sytems.  This leads to the composite S0 spectrum having
both stronger H$\delta_A$ absorption and lower $D_N(4000)$ than the
composite spectrum of the bright, red ellipticals.  Because
H$\delta_A$ is a more sensitive tracer of relative age than
$D_N(4000)$ in the range of mean stellar ages probed here ($1-3$ Gyr;
K03), we use H$\delta_A$ and estimate an age difference of $\sim0.5$
Gyr.  Thus even the composite spectrum of the bright, red S0s
corresponds to a younger mean age than that of the bright, red
ellipticals.

In concert, these results indicate that on average, the S0s are
younger (by $\sim0.5-1$ Gyr) and less homogeneous in age than the
ellipticals; this age difference is consistent with the measured color
offset between S0 and elliptical galaxies in a cluster at $z=1.106$
\citep[$\Delta t\sim1$ Gyr;][]{mei:06}.  However, we note that even a
difference in age of 1 Gyr at $z=0.8$ will have faded and not be
detectable by the current epoch, $i.e.$ 7 Gyr later.

Although separating ellipticals from S0s is challenging and there is
usually considerable contamination between the two types, we find that
the ellipticals and S0s in MS1054 are not drawn from the same parent
population: while many of the S0s are as old as the ellipticals, there
are a number of S0s that are blue (18\%) and/or have a post-starburst
signature (21\%).  Our results confirm that an average elliptical in
MS1054 is younger than an average S0, and that it is possible to
reliably separate ellipticals and S0s at $z=0.83$ with high resolution
imaging ($\sim0.4$\hi kpc).

\subsection{Red Mergers}

For completeness, we include a brief analysis of the members that are
in red, merging systems.  MS1054's unusually high fraction of red
galaxy-galaxy mergers ($\sim17$\%) was first reported by
\citet{vandokkum:99} and later spectroscopically confirmed by
\citet{tran:05a}.  The composite spectrum of the 17 red merging
members from \citet{tran:05a} is included in Fig.~\ref{coadd} (bottom),
and the corresponding spectral indices listed in Table~\ref{indices}.
The red merger spectrum is essentially identical to both the composite
absorption-line and elliptical spectrum, and it has similar values for
H$\delta_A$ and $D_N(4000)$.  These spectral diagnostics reinforce our
previous conclusions that the galaxies in these red, merging systems
are dominated by an old stellar population with no ongoing star
formation, and that most of the massive galaxies in MS1054 evolved
from dissipationless mergers.

\subsection{H$\delta_A$ vs. $D_N(4000)$:  Comparing Star Formation Histories} 

\begin{figure}
\plotone{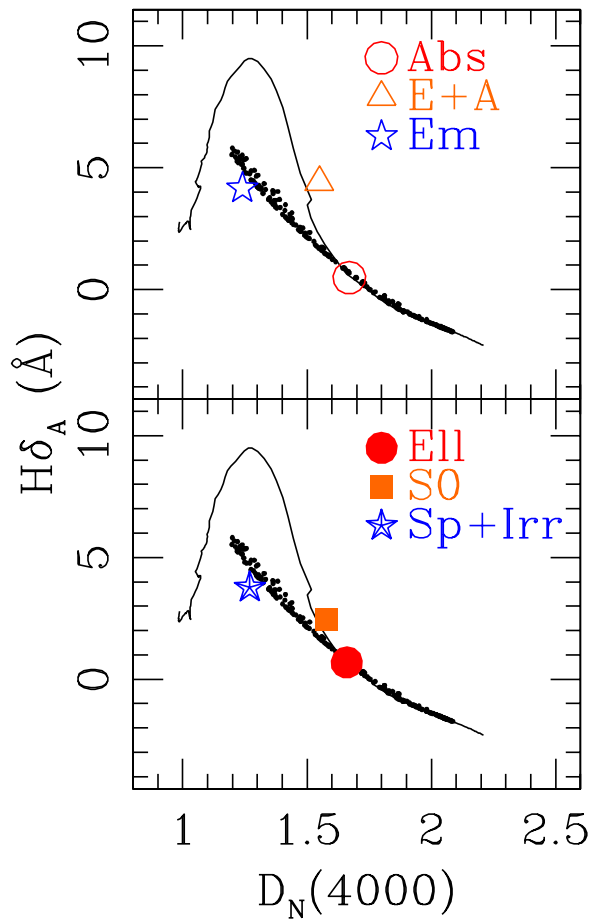}
\caption{Using our composite spectra, we measure the spectral indices
H$\delta_A$ and $D_N(4000)$ to compare mean stellar ages.  The
different spectral and morphological classes are shown in the top and
middle panels respectively; errorbars are smaller than the symbol
sizes (see Table~\ref{indices}).  The red mergers lie at the same
point as the absorption-line members.  The solar metallicity BC03
models from \citet{kauffmann:03} for continuous star formation (small
dots) and single starburst (curve) are included in each panel.  The
spectral diagnostics show that the composite S0 spectrum has a younger
mean stellar age than the composite elliptical.
\label{Hd_D4000}}
\end{figure}

In Fig.~\ref{Hd_D4000}, we plot H$\delta_A-D_N(4000)$ for the
different spectral and morphological types; the figure also includes
the models, both for mixed star formation histories (dots) and a
single burst (curve), from K03.  The mixed star formation histories
include combinations of continuous star formation with random bursts
of varying strength.  To zeroeth order, $D_N(4000)$ traces the
mass-to-light ratio such that galaxies with larger $(M/L)$ ratios,
$i.e.$ those dominated by older stars, have higher $D_N(4000)$
indices.  However, at fixed $D_N(4000)$, galaxies with stronger
H$\delta_A$ absorption have a higher fraction of young stars and
therefore smaller $(M/L)$ ratios.  Combining $D_N(4000)$ with
H$\delta_A$ thus enables us to better gauge whether a galaxy has
experienced a burst within the last 2 Gyr, and whether this recent
episode increased the stellar mass significantly.

Immediately apparent from the top panel of Fig.~\ref{Hd_D4000} is that
the recent ($<1.5$ Gyr) star formation histories of the
absorption-line, post-starburst, and emission-line members are very
different.  Assuming solar metallicity, we find that the indices for
the composite absorption-line spectrum are consistent with a single
starburst that is $\sim2.8$ Gyr old ($z_f\sim1.8$).  The indices for
the composite post-starburst spectrum are also consistent with a
single starburst, but one that is only $\sim1.6$ Gyr old
($z_f\sim1.2$).  In comparison, the indices for the emission-line
spectrum differ from a single starburst and indicate a continuous
and/or bursting star formation history (see K03) where the mean
stellar age is $\sim0.4$ Gyr ($z_f\sim0.9$).

Fig.~\ref{Hd_D4000} also shows H$\delta_A-D_N(4000)$ for the
morphological types (bottom panel).  The composite elliptical spectrum
has H$\delta_A$ and $D_N(4000)$ values comparable to those of the
absorption-line spectrum.  This mirrors the result from their colors
that the elliptical and absorption-line galaxies are the oldest
members (see \S\ref{cmr}).  In comparison, the S0s have stronger
H$\delta_A$ absorption and lower $D_N(4000)$, consistent with our
conclusion that the S0s have had more recent star formation and are
not as uniformly old as the absorption-line members.

The H$\delta_A-D_N(4000)$ values of the composite absorption-line
spectrum confirm that these are the oldest galaxies in MS1054, a
result that is consistent with their small scatter in color (see
\S\ref{cmr}); only stellar populations (solar metallicity) with mean
ages $>2.5$ Gyr can have $D_N(4000)>1.65$ and
$H\delta_A\lesssim0.5$\AA~(K03).  Using H$\delta_A-D_N(4000)$
reinforces our conclusion that selecting by spectral type alone is
most effective at isolating a homogeneously old population.


\section{Signs of Age on the Red Sequence:  \\
Galaxy Color \& H$\delta$ Absorption}\label{age}

With our spectral survey, we are able to identify a uniformly old
population of galaxies in MS1054: in terms of both their spectral and
photometric properties, the absorption-line members are older and more
homogeneous in mean stellar age than the post-starburst,
emission-line, and morphologically classified E+S0 members.  The
absorption-line members make up 63\% of the current cluster
population, span a wide range in luminosity (factor of 25), and have a
small scatter in color.  Spectral diagnostics show that the composite
absorption-line member has a mean stellar age of $\sim2.8$ Gyr
($z_f\sim1.8$).

However, many of the cluster galaxies continue to evolve in color,
morphology, and spectral line strength, $i.e.$ the red sequence
continues to assemble.  A significant fraction ($\sim15$\%) of
MS1054's members are post-starburst systems that are likely to evolve
into absorption-line members, and there are a number of
spectroscopically confirmed red mergers that can only evolve into
massive, passive members \citep{vandokkum:99,tran:05a}.  In the
following, we compare different subsets of the cluster galaxies to
further illustrate how the red sequence in MS1054 has evolved since
$z\sim1.8$, and how it continues to be populated at $z<0.83$.

\subsection{A Correlation Between H$\delta_A$ \& Color Offset}

To test for variations in relative age between the absorption-line
population and other members on the red sequence, $e.g.$ the E+As, we
consider how the H$\delta_A$ equivalent width varies as a function of
offset in $(V-i_{775})$ for the cluster galaxies.  Because there may
be a trend in H$\delta_A$ with magnitude due to, $e.g.$ the
metallicity-luminosity relation \citep{gallazzi:06}, we normalize
H$\delta_A$ by subtracting from the measured H$\delta$ EW the
H$\delta_A-i_{775}$ relation where the latter is determined by a
least-squares fit to the red absorption-line members:

\begin{equation}
H\delta_A=0.047-1.670(i_{775}-22)
\end{equation}

\noindent We note that this trend between H$\delta_A$ and $i_{775}$ is
not statistically significant using a Spearman rank test.  However, we
use the normalized H$\delta$, hereafter referred to as $\Delta
H\delta_A$, to ensure the robustness of our results.

\begin{figure}
\plotone{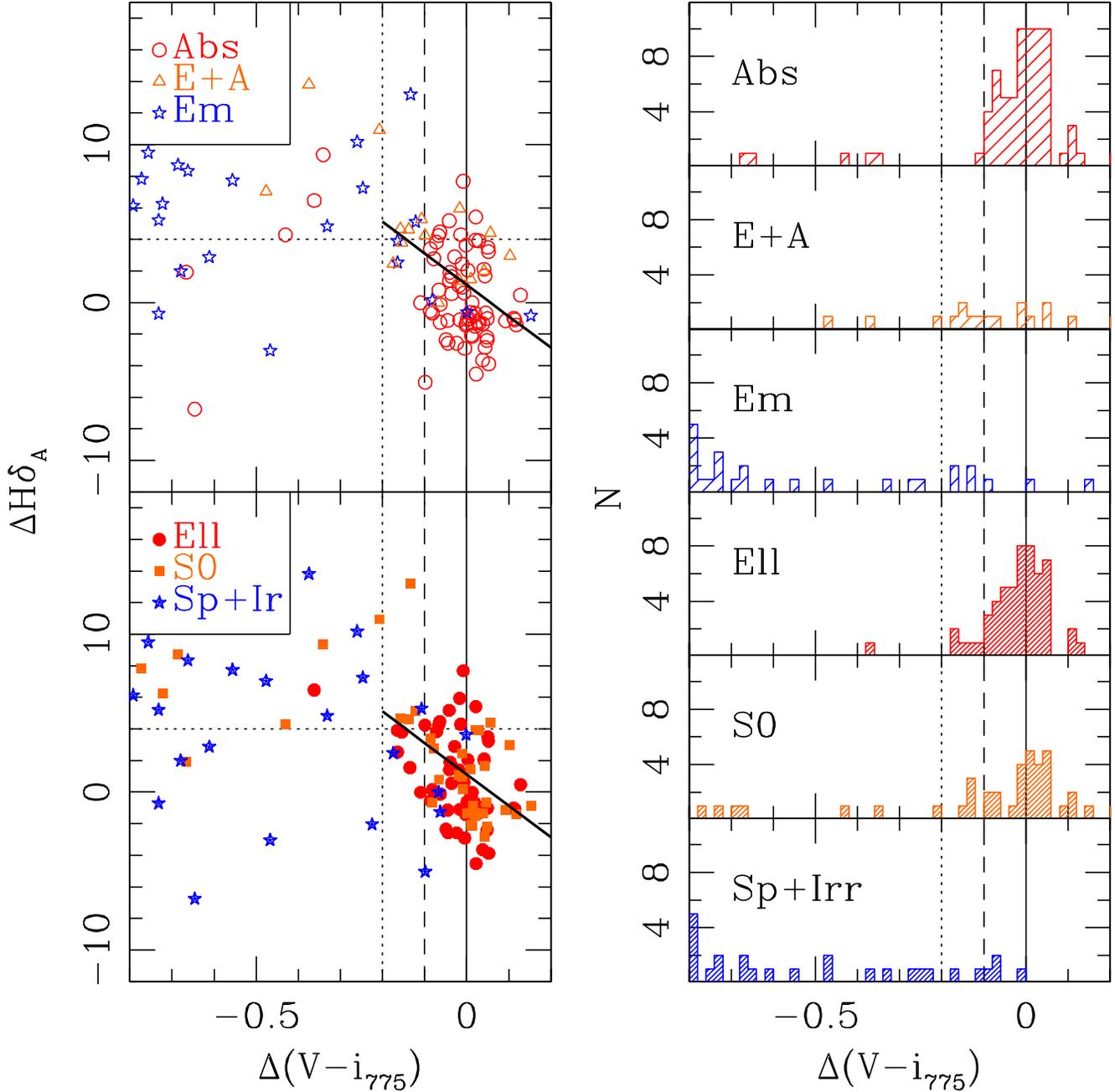}
\caption{{\it Left Panels:} Normalized H$\delta_A$ equivalent width
versus offset in $(V-i_{775})$; for clarity, we do not show the full
range of values (see Table~1).  The dotted horizontal line corresponds
to H$\delta_A=4$\AA, the vertical dotted line at
$\Delta(V-i_{775})=-0.2$ divides blue and red members, and the dashed
line at $\Delta(V-i_{775})=-0.1$ highlights the tight color
distribution of the absorption-line members.  Remarkably, 8/10 of the
members in the adjacent color bin, $-0.2\leq\Delta(V-i_{775})<-0.1$,
show post-starburst signatures.  We find a trend of decreasing
H$\delta$ absorption with color for the red members (heavy solid
diagonal line; $>95$\% confidence). {\it Right Panels:} Histograms of
$\Delta(V-i_{775})$ for the same galaxy classes shown in the left
panels; the vertical lines correspond to the same $\Delta(V-i_{775})$
references.
\label{dHd_dVi}}
\end{figure}

Fig.~\ref{dHd_dVi} (right panels) highlights the impressively tight
color distribution of the absorption-line population: 92\% (66/72) of
the absorption-line members have $\Delta(V-i_{775})>-0.1$.  The
ellipticals also show a tight color distribution with 89\% (49/55)
having $\Delta(V-i_{775})>-0.1$.  In comparison, only 72\% (28/39) of
the S0s are as red.

Fig.~\ref{dHd_dVi} (left panels) reveals that even on the red
sequence, there is a trend between color deviation and H$\delta_A$
absorption.  We fit a least squares to $\Delta$H$\delta_A$ vs.
$\Delta(V-i_{775})$ for all of the members with
$-0.2\leq\Delta(V-i_{775})\leq0.2$ and $\sigma_{1D}>150$\kms; the lower limit
on the internal velocity dispersion provides a mass-limited sample and
ensures that we are not introducing a luminosity bias 
into the fit.  With these 52 members, we find that:

\begin{equation}
\Delta H\delta_A = 1.13-19.93 \Delta(V-i_{775})
\end{equation}

\noindent The Spearman rank test \citep{press:92} confirms that the
trend of weakening H$\delta$ absorption with redder colors is
significant with $>95$\% confidence.  In terms of how large the age
variations are for such a relatively small difference in color, we
note that a member with $(V-i_{775})=1.6$ is older than a member with
$(V-i_{775})=1.5$ by $\sim1$ Gyr; this is determined by comparing the
corresponding $(U-B)_z$ values of 0.35 and 0.27 to a BC03 single burst
model (solar metallicity; B06).


Assuming that the observed trend between weakening H$\delta_A$
absorption and redder colors is due to differences in mean steller age
implies that even amongst the absorption-line members, there is a
trend in age with color.  To test whether this is the case, we
separate the red absorption-line members into two smaller bins that
straddle $\Delta(V-i_{775})=0$ [bin size of $\Delta(V-i_{775})=0.1$].
We generate composite spectra for these two subsets and measure their
spectral indices (Table~\ref{indices}).

The composite spectrum for the absorption-line members that are redder
than the CM relation [$0\leq\Delta(V-i_{775})\leq0.1$; 36 members] has
an H$\delta_A$ EW of 0.1\AA~(essentially none) and $D_N(4000)=1.73$
(the highest value for any composite spectrum).  If we use color,
H$\delta_A$, and $D_N(4000)$ as proxies for mean stellar age, this
would imply that 1) the absorption-line members with colors that are
redder than the fitted CM relation are the oldest members and 2) any
absorption-line members with more recent star formation will
consequently be bluer and have stronger H$\delta_A$ absorption as well
as a lower $D_N(4000)$ index.  We see this is exactly the case in
MS1054: the composite spectrum for the absorption-line members that
are bluer than the CM relation [$-0.1\leq\Delta(V-i_{775})<0$; 25
members] has a measurably stronger H$\delta_A$ EW of $1.0$\AA~and
lower $D_N(4000)$ of 1.66.  Equally remarkable is that the adjacent
color bin [$-0.2\leq\Delta(V-i_{775})<-0.1$] is dominated by
post-starburst members and thus has even stronger H$\delta_A$
absorption (Fig.~\ref{dHd_dVi}).

\subsection{Explaining the Correlation:  Age vs. Metallicity}

While the correlation between redder colors and weakening H$\delta_A$
absorption can result from variations in either age or metallicity
\citep{terlevich:99}, our observations indicate that age is the
underlying cause because: 1) The $i_{775}$ magnitude distributions of
the absorption-line members in the two color bins are
indistinguishable (K-S test), thus we are not inadvertently dividing
the members into high/low metallicity groups corresponding to
bright/faint members; 2) As a precaution, we have already removed any
possible trend of H$\delta_A$ with magnitude by subtracting the
H$\delta_A-i_{775}$ relation from the measured value; and 3) Most
compelling is that of the 10 members with
$-0.2\leq\Delta(V-i_{775})<-0.1$, six are post-starburst systems and
two have strong Balmer absorption and only weak (O[II]$\sim5-7$\AA)
emission, meaning they too experienced a recent starburst.  If the
trend between H$\delta_A$ and color is due to metallicity rather than
age, one would not expect this color bin to be dominated by
post-starburst systems, $i.e.$ galaxies that definitely have younger
mean stellar ages.

To further test our hypothesis that the observed trend between
H$\delta_A$ and color offset is due to variations in age, we compare
our results to two simple models: Case 1) assuming a constant age of 3
Gyr and allowing the metallicity to change from $Z=Z_{\odot}$ at
$B_z=20$ to $Z=0.2Z_{\odot}$ at $B_z=24$; and Case 2) assuming
constant solar metallicity and allowing the age to change from 3 Gyr
at $B_z=20$ to 1.4 Gyr at $B_z=24$.  We use the BC03 models with the
Padova 1994 evolutionary tracks and the Chabrier initial mass function
to fit the following relations for Case 1:

\begin{equation}
H\delta_A=-0.075 + 0.640 (B_z-20)
\end{equation}
\begin{equation}
\Delta(V-i_{775})=-0.055(B_z-20)
\end{equation}

\noindent and for Case 2:

\begin{equation}
H\delta_A=-0.075 + 0.959 (B_z-20)
\end{equation}
\begin{equation}
\Delta(V-i_{775})=-0.047(B_z-20)
\end{equation}

\begin{figure}
\plotone{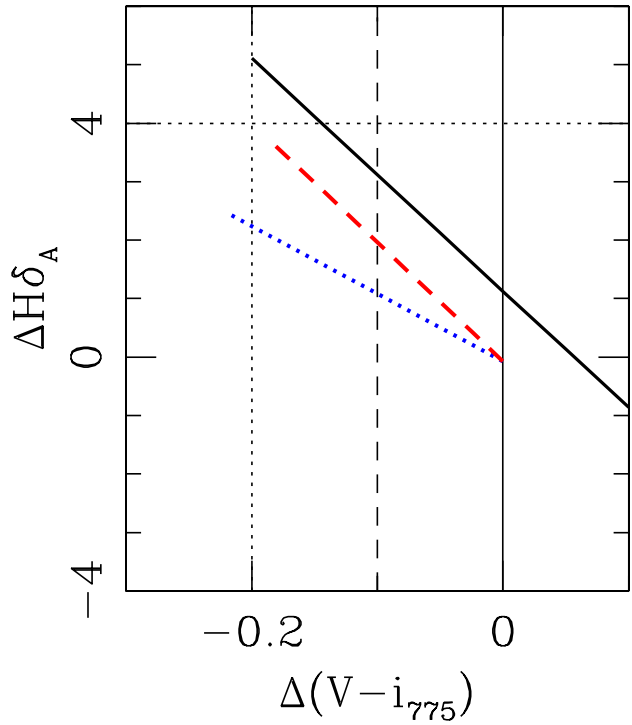}
\caption{To determine whether the measured trend between H$\delta_A$
  and $\Delta(V-i_{775})$ is due to age or metallicity, we compare to
  the BC03 single-burst models.  The vertical and horizontal lines are
  as in Fig.~\ref{dHd_dVi} (left panels), and the heavy diagonal is
  the fit to the red members.  In Case 1 (dotted diagonal), we assume
  galaxies with constant age (3 Gyr) and changing metallicity
  ($0.2Z_{\odot}-Z_{\odot}$), and in Case 2 (dashed diagonal), we
  assume constant solar metallicity and ages of $1.4-3$ Gyr.  The
  measured fit is best matched by the trend due to age variations
  alone (Case 2; dashed diagonal).}
\label{degmodel}
\end{figure}

Figure~\ref{degmodel} shows how the H$\delta_A$ absorption and color
offset vary for the two models.  Both models show similar trends of
weakening H$\delta_A$ absorption with redder colors.  However, the
observed trend is best matched by Case 2 where metallicity is constant
($Z=Z_{\odot}$) and age varies ($1.4-3$ Gyr).  For Case 1, it is
possible to steepen the H$\delta_A-\Delta(V-i_{775})$ trend to better
match the data, but this would require metallicities significantly
lower ($\lesssim0.2Z_{\odot}$) than that measured for early-type
galaxies in the local universe
\citep[$Z=0.8-1.6Z_{\odot}$;][]{gallazzi:06}.

We have presented several lines of evidence supporting a trend between
H$\delta_A$ and color offset that is due to age variations alone.
However, we acknowledge that we cannot exclude metallicity as the
underlying cause, either in part or in full, at a statistically robust
level.  For example, the least-squares fit defined by the reddest
[$\Delta(V-i_{775})\geq-0.1$] members has a slightly flatter slope
($m=-16.94$) and thus does not deviate as strongly from Case 1.  Only
by measuring the spectral indices needed to disentangle age and
metallicity can this issue be fully resolved; such an analysis is
beyond the scope of this paper.

\subsection{Estimating the Age Variations on the Red Sequence}

We conclude that the color scatter in MS1054's red sequence is driven
by differences in mean stellar age.  To estimate the age difference
bracketed by the oldest and youngest average red sequence member at
$z=0.83$, we use the H$\delta_A$ and $D_N(4000)$ values from the
composite spectra of the reddest absorption-line members and the E+As
(Table~\ref{indices}).  For the $\Delta(V-i_{775})>0$ absorption-line
spectrum, comparing both diagnostics to the single burst model (K03)
shows that they have a mean age of $\sim3$ Gyr while the E+As have a
mean age of $\sim1.6$ Gyr.  Assuming no future star formation, the
current population of E+As will evolve into absorption-line members
that lie on the red sequence in $<2$ Gyr \citep{barger:96}.  Both
\citet{vandokkum:98a} and \citet{bower:98} confirm using mixed star
formation models that color scatter decreases rapidly as long as most
of the members ($>75$\%) formed at higher redshifts ($z\gtrsim2$).
Considering the large lookback time to $z=0.83$ (7 Gyr), the larger
color scatter introduced by the E+As will decrease and be
indistinguishable from that of the absorption-line members by the
present epoch.

\section{Summary \& Conclusions}

We present an extensive spectroscopic survey of MS~1054--03 using
observations collected with Keck/LRIS.  From our magnitude limited
survey, we isolate 129 ($Q_z=3$) cluster galaxies and measure MS1054's
redshift and velocity dispersion to be $z=0.8307\pm0.0004$ and
$\sigma_z=1156\pm82$\kms.  We estimate MS1054's virial radius and mass
to be $1.8\pm0.1$\hi~Mpc and $3.3\pm0.6\times10^{15}M_{\odot}$, but note
that the cluster has a significant amount of substructure.  The
absorption-line, post-starburst (E+A), and emission-line galaxies make
up $63\pm7$\%, $15\pm4$\%, and $23\pm4$\% of the cluster population,
respectively.

Combining our spectroscopy with magnitudes, colors, and Hubble types
from HST/ACS imaging \citep{postman:05,blakeslee:06}, we compare the
different galaxy classes.  In the color-magnitude diagram, the
absorption-line members define a tight red sequence over a span of
$\sim3.5$ magnitudes: their intrinsic scatter in $(V-i_{775})$ color
is only $0.048\pm0.009$, corresponding to a scatter in Johnson
$(U-B)_z$ of $0.041$.  The color scatter for the absorption-line
members is comparable to that of the morphologically classified
ellipticals ($\sigma_{Vi}=0.055\pm0.008$), but it is measurably
smaller than that of the combined E+S0 sample
($\sigma_{Vi}=0.072\pm0.010$).  Our survey demonstrates that
spectroscopy is extremely effective at identifying a homogeneously old
population over a wide range in luminosity ($L\gtrsim0.5L^{\ast}$),
especially at $z\gtrsim0.8$ where visually separating ellipticals and
S0s becomes quite challenging.

In MS1054, the scatter in $(U-V)_z$ for both the
absorption-line and elliptical members is more than twice that of the
ellipticals in Coma: $\sigma_{UVz}=0.091\pm0.009$
vs. $\sigma_{UVz}=0.036\pm0.020$.  The difference in the color scatter
is consistent with passive evolution of a population where most of the
absorption-line members ($>75$\%) formed by $z\sim2$
\citep{vandokkum:98a,bower:98}, and all of them by $z\sim1.2$.  The
absorption-line galaxies in MS1054 are likely to evolve into the
oldest members of local clusters such as Coma.

Our analysis indicates that the color scatter on the red sequence in
MS1054 is due to differences in mean stellar age.  Considering only
red [$\Delta(V-i_{775})>-0.2$] members, we find a trend ($>95$\%
confidence) between weakening H$\delta$ absorption and redder colors
that we conclude is due to age.  We estimate that red sequence members
differ in mean stellar age by up to $\sim1.5$ Gyr.  However, such a
difference in mean age and the resulting increase in color scatter are
both negligible by the present epoch given the lookback time to
$z=0.83$ of $\sim7$ Gyr.

The current generation of transitional members, $e.g.$ the E+As and
red mergers, include luminous/massive galaxies that will populate the
red sequence across a wide range in luminosity.  However, assuming
that the red sequence continues to acquire more members at $z<0.83$ as
blue members fade and redden, we note that these later additions to
the red sequence will be mostly faint ($L<L^{\ast}$), low-mass
($\sigma_{1D}<100$\kms) members.  The current deficit of
morphologically classified ellipticals and S0s in particular suggests
that most of these later additions to the red sequence will evolve
into S0s \citep{tran:05a,postman:05,holden:06}.  This would result in
a difference in mean stellar age between the bright ellipticals and
the faint S0s, but such a difference is observed in lower redshift
clusters, $e.g.$ Coma \citep{poggianti:01b}.

Lastly, we generate composite spectra for the different galaxy classes
and measure their spectral indices, $e.g.$ H$\delta_A$ and
$D_N(4000)$.  Both color and spectral indices confirm that the
composite absorption-line spectrum has the oldest mean stellar
population ($\sim3$ Gyr), and that the composite elliptical spectrum
is equally as old.  Remarkably, the spectral diagnostics also indicate
that the composite S0 spectrum is $\sim0.5-1$ Gyr younger than the
composite elliptical spectrum.  This is due to the fact that while
most of the S0s are as red (old) as the ellipticals, $\sim18$\% are
blue and $\sim21$\% are also post-starburst systems.  The difference
in mean stellar age between an average elliptical and S0 in MS1054 is
consistent with the age difference implied by the color offset
observed between ellipticals and S0s in a cluster at $z=1.106$
\citep{mei:06}.  Although there is likely to be some mixing 
when separating ellipticals and S0s, our results show that
at least in MS1054 at $z=0.83$, the average elliptical is measurably
older than the average S0 member.

\acknowledgments

It is a pleasure to thank G. Kauffmann for providing her star
formation models, M. Jee for his weak-lensing map of MS1054, and
I. Gioia for her X-ray map of MS1054.  K. Tran acknowledges support
from the NSF Astronomy \& Astrophysics Postdoctoral fellowship under
award AST-0502156 and from the NOVA fellowship program; she also
thanks S. Trager for technical help with the star formation models.
ACS was developed under NASA contract NAS5-32865, and this research
was supported under NASA grant NAG5-7697.  Finally, the authors extend
special thanks to those of Hawaiian ancestry on whose sacred mountain
we are privileged to be guests.

\bibliographystyle{/Users/vy/aastex/apj}
\bibliography{/Users/vy/aastex/tran}

\begin{thebibliography}{71}
\expandafter\ifx\csname natexlab\endcsname\relax\def\natexlab#1{#1}\fi

\bibitem[{{Balogh} {et~al.}(1999){Balogh}, {Morris}, {Yee}, {Carlberg}, \&
  {Ellingson}}]{balogh:99}
{Balogh}, M.~L., {Morris}, S.~L., {Yee}, H.~K.~C., {Carlberg}, R.~G., \&
  {Ellingson}, E. 1999, \apj, 527, 54

\bibitem[{{Barger} {et~al.}(1996){Barger}, {Aragon-Salamanca}, {Ellis},
  {Couch}, {Smail}, \& {Sharples}}]{barger:96}
{Barger}, A.~J., {Aragon-Salamanca}, A., {Ellis}, R.~S., {Couch}, W.~J.,
  {Smail}, I., \& {Sharples}, R.~M. 1996, \mnras, 279, 1

\bibitem[{{Baugh} {et~al.}(1996){Baugh}, {Cole}, \& {Frenk}}]{baugh:96}
{Baugh}, C.~M., {Cole}, S., \& {Frenk}, C.~S. 1996, \mnras, 283, 1361

\bibitem[{Beers {et~al.}(1990)Beers, Flynn, \& Gebhardt}]{beers:90}
Beers, T.~C., Flynn, K., \& Gebhardt, K. 1990, \aj, 100, 32

\bibitem[{Bertin \& Arnouts(1996)}]{bertin:96}
Bertin, E. \& Arnouts, S. 1996, \aap, 117, 393

\bibitem[{{Blakeslee} {et~al.}(2006){Blakeslee}, {Holden}, {Franx}, {Rosati},
  \& et~al.}]{blakeslee:06}
{Blakeslee}, J.~P., {Holden}, B.~P., {Franx}, M., {Rosati}, P., \& et~al. 2006,
  \apj

\bibitem[{{Bower} {et~al.}(1998){Bower}, {Kodama}, \& {Terlevich}}]{bower:98}
{Bower}, R.~G., {Kodama}, T., \& {Terlevich}, A. 1998, \mnras, 299, 1193

\bibitem[{{Bower} {et~al.}(1992){Bower}, {Lucey}, \& {Ellis}}]{bower:92}
{Bower}, R.~G., {Lucey}, J.~R., \& {Ellis}, R.~S. 1992, \mnras, 254, 601+

\bibitem[{{Bruzual} \& {Charlot}(2003)}]{bruzual:03}
{Bruzual}, G. \& {Charlot}, S. 2003, \mnras, 344, 1000

\bibitem[{{Butcher} \& {Oemler}(1984)}]{butcher:84}
{Butcher}, H. \& {Oemler}, A. 1984, \apj, 285, 426

\bibitem[{{Couch} \& {Sharples}(1987)}]{couch:87}
{Couch}, W.~J. \& {Sharples}, R.~M. 1987, \mnras, 229, 423

\bibitem[{{Diaferio} {et~al.}(2001){Diaferio}, {Kauffmann}, {Balogh}, {White},
  {Schade}, \& {Ellingson}}]{diaferio:01}
{Diaferio}, A., {Kauffmann}, G., {Balogh}, M.~L., {White}, S.~D.~M., {Schade},
  D., \& {Ellingson}, E. 2001, \mnras, 323, 999

\bibitem[{Donahue {et~al.}(1998)Donahue, Voit, Gioia, Luppino, Hughes, \&
  Stocke}]{donahue:98}
Donahue, D., Voit, G., Gioia, I., Luppino, G., Hughes, J.~P., \& Stocke, J.~T.
  1998, \apj, 502, 550

\bibitem[{{Dressler} \& {Gunn}(1983)}]{dressler:83}
{Dressler}, A. \& {Gunn}, J.~E. 1983, \apj, 270, 7

\bibitem[{{Dressler} {et~al.}(1997){Dressler}, {Oemler}, {Couch}, {Smail},
  {Ellis}, {Barger}, {Butcher}, {Poggianti}, \& {Sharples}}]{dressler:97}
{Dressler}, A., {Oemler}, A.~J., {Couch}, W.~J., {Smail}, I., {Ellis}, R.~S.,
  {Barger}, A., {Butcher}, H., {Poggianti}, B.~M., \& {Sharples}, R.~M. 1997,
  \apj, 490, 577

\bibitem[{{Dressler} {et~al.}(2004){Dressler}, {Oemler}, {Poggianti}, {Smail},
  {Trager}, {Shectman}, {Couch}, \& {Ellis}}]{dressler:04}
{Dressler}, A., {Oemler}, A.~J., {Poggianti}, B.~M., {Smail}, I., {Trager}, S.,
  {Shectman}, S.~A., {Couch}, W.~J., \& {Ellis}, R.~S. 2004, \apj, 617, 867

\bibitem[{{Dressler} \& {Shectman}(1988)}]{dressler:88}
{Dressler}, A. \& {Shectman}, S.~A. 1988, \aj, 95, 985

\bibitem[{{Dressler} {et~al.}(1999){Dressler}, {Smail}, {Poggianti}, {Butcher},
  {Couch}, {Ellis}, \& {Oemler}}]{dressler:99}
{Dressler}, A., {Smail}, I., {Poggianti}, B.~M., {Butcher}, H., {Couch}, W.~J.,
  {Ellis}, R.~S., \& {Oemler}, A.~J. 1999, \apjs, 122, 51

\bibitem[{{Ellingson} {et~al.}(2001){Ellingson}, {Lin}, {Yee}, \&
  {Carlberg}}]{ellingson:01}
{Ellingson}, E., {Lin}, H., {Yee}, H.~K.~C., \& {Carlberg}, R.~G. 2001, \apj,
  547, 609

\bibitem[{{Ellis} {et~al.}(1997){Ellis}, {Smail}, {Dressler}, {Couch},
  {Oemler}, {Butcher}, \& {Sharples}}]{ellis:97}
{Ellis}, R.~S., {Smail}, I., {Dressler}, A., {Couch}, W.~J., {Oemler}, A.~J.,
  {Butcher}, H., \& {Sharples}, R.~M. 1997, \apj, 483, 582

\bibitem[{{Fisher} {et~al.}(1998){Fisher}, {Fabricant}, {Franx}, \& {van
  Dokkum}}]{fisher:98}
{Fisher}, D., {Fabricant}, D., {Franx}, M., \& {van Dokkum}, P. 1998, \apj,
  498, 195+

\bibitem[{{Ford} {et~al.}(1998){Ford}, {Bartko}, {Bely}, {Broadhurst},
  {Burrows}, {Cheng}, {Clampin}, \& {et al.}}]{ford:98}
{Ford}, H.~C., {Bartko}, F., {Bely}, P.~Y., {Broadhurst}, T., {Burrows}, C.~J.,
  {Cheng}, E.~S., {Clampin}, M., \& {et al.} 1998, in Proc. SPIE Vol. 3356, p.
  234-248, Space Telescopes and Instruments V, Pierre Y. Bely; James B.
  Breckinridge; Eds., ed. P.~Y. {Bely} \& J.~B. {Breckinridge}, 234--248

\bibitem[{{Gallazzi} {et~al.}(2006){Gallazzi}, {Charlot}, {Brinchmann}, \&
  {White}}]{gallazzi:06}
{Gallazzi}, A., {Charlot}, S., {Brinchmann}, J., \& {White}, S.~D.~M. 2006,
  \mnras, 370, 1106

\bibitem[{{Gioia} {et~al.}(2004){Gioia}, {Braito}, {Branchesi}, {Della Ceca},
  {Maccacaro}, \& {Tran}}]{gioia:04}
{Gioia}, I.~M., {Braito}, V., {Branchesi}, M., {Della Ceca}, R., {Maccacaro},
  T., \& {Tran}, K.-V. 2004, \aap, 419, 517

\bibitem[{Gioia \& Luppino(1994)}]{gioia:94}
Gioia, I.~M. \& Luppino, G.~A. 1994, \apjs, 94, 583

\bibitem[{{Goto} {et~al.}(2005){Goto}, {Postman}, {Cross}, {Illingworth},
  {Tran}, {Magee}, \& {Franx}}]{goto:05}
{Goto}, T., {Postman}, M., {Cross}, N.~J.~G., {Illingworth}, G.~D., {Tran}, K.,
  {Magee}, D., \& {Franx}, M. 2005, \apj, 621, 188

\bibitem[{{Gunn} \& {Stryker}(1983)}]{gunn:83}
{Gunn}, J.~E. \& {Stryker}, L.~L. 1983, \apjs, 52, 121

\bibitem[{{Hoekstra} {et~al.}(2000){Hoekstra}, {Franx}, \&
  {Kuijken}}]{hoekstra:00}
{Hoekstra}, H., {Franx}, M., \& {Kuijken}, K. 2000, \apj, 532, 88

\bibitem[{{Holden} {et~al.}(2006){Holden}, {Franx}, {Illingworth}, {Postman},
  {Blakeslee}, {Homeier}, {Demarco}, {Ford}, {Rosati}, {Kelson}, \&
  {Tran}}]{holden:06}
{Holden}, B.~P., {Franx}, M., {Illingworth}, G.~D., {Postman}, M., {Blakeslee},
  J.~P., {Homeier}, N., {Demarco}, R., {Ford}, H.~C., {Rosati}, P., {Kelson},
  D.~D., \& {Tran}, K.-V.~H. 2006, \apjl, 642, L123

\bibitem[{{Homeier} {et~al.}(2005){Homeier}, {Demarco}, {Rosati}, {Postman},
  {Blakeslee}, {Bouwens}, {Bradley}, \& many~more authors}]{homeier:05}
{Homeier}, N.~L., {Demarco}, R., {Rosati}, P., {Postman}, M., {Blakeslee},
  J.~P., {Bouwens}, R.~J., {Bradley}, L.~D., \& many~more authors. 2005, \apj,
  621, 651

\bibitem[{{Jee} {et~al.}(2005){Jee}, {White}, {Ford}, {Blakeslee},
  {Illingworth}, {Coe}, \& {Tran}}]{jee:05}
{Jee}, M.~J., {White}, R.~L., {Ford}, H.~C., {Blakeslee}, J.~P., {Illingworth},
  G.~D., {Coe}, D.~A., \& {Tran}, K.-V.~H. 2005, \apj, 634, 813

\bibitem[{{Jeltema} {et~al.}(2001){Jeltema}, {Canizares}, {Bautz}, {Malm},
  {Donahue}, \& {Garmire}}]{jeltema:01}
{Jeltema}, T.~E., {Canizares}, C.~R., {Bautz}, M.~W., {Malm}, M.~R., {Donahue},
  M., \& {Garmire}, G.~P. 2001, \apj, 562, 124

\bibitem[{{J{\o}rgensen} {et~al.}(1999){J{\o}rgensen}, {Franx}, {Hjorth}, \&
  {van Dokkum}}]{jorgensen:99}
{J{\o}rgensen}, I., {Franx}, M., {Hjorth}, J., \& {van Dokkum}, P.~G. 1999,
  \mnras, 308, 833

\bibitem[{{Kauffmann}(1995)}]{kauffmann:95}
{Kauffmann}, G. 1995, \mnras, 274, 153

\bibitem[{{Kauffmann} {et~al.}(2003){Kauffmann}, {Heckman}, {White}, {Charlot},
  \& {et al.}}]{kauffmann:03}
{Kauffmann}, G., {Heckman}, T.~M., {White}, S.~D.~M., {Charlot}, S., \& {et
  al.} 2003, \mnras, 341, 33

\bibitem[{Kelson(1998)}]{kelson:98}
Kelson, D.~D. 1998, Ph.D. thesis, University of California at Santa Cruz (Santa
  Cruz, CA: University of California)

\bibitem[{{Kelson} {et~al.}(2000){Kelson}, {Illingworth}, {van Dokkum}, \&
  {Franx}}]{kelson:00c}
{Kelson}, D.~D., {Illingworth}, G.~D., {van Dokkum}, P.~G., \& {Franx}, M.
  2000, \apj, 531, 184

\bibitem[{{Kodama} {et~al.}(1998){Kodama}, {Arimoto}, {Barger}, \&
  {Arag'on-Salamanca}}]{kodama:98}
{Kodama}, T., {Arimoto}, N., {Barger}, A.~J., \& {Arag'on-Salamanca}, A. 1998,
  \aap, 334, 99

\bibitem[{Kurtz {et~al.}(1992)Kurtz, Mink, Wyatt, Fabricant, Torres, Kriss, \&
  Tonry}]{kurtz:92}
Kurtz, M.~J., Mink, D.~J., Wyatt, W.~F., Fabricant, D.~G., Torres, G., Kriss,
  G.~A., \& Tonry, J.~L. 1992, in Astronomical Data Analysis Software and
  Systems I, ed. d.~M. Worrall, C.~Biemesderfer, \& J.~Barnes, Vol.~25 (A.S.P.
  Conference Series), 432

\bibitem[{{Lubin} {et~al.}(2002){Lubin}, {Oke}, \& {Postman}}]{lubin:02}
{Lubin}, L.~M., {Oke}, J.~B., \& {Postman}, M. 2002, \aj, 124, 1905

\bibitem[{{Mei} {et~al.}(2006){Mei}, {Blakeslee}, {Stanford}, {Holden},
  {Rosati}, {Strazzullo}, {Homeier}, \& {et al.}}]{mei:06}
{Mei}, S., {Blakeslee}, J.~P., {Stanford}, S.~A., {Holden}, B.~P., {Rosati},
  P., {Strazzullo}, V., {Homeier}, N., \& {et al.} 2006, \apj, 639, 81

\bibitem[{{Newberry} {et~al.}(1990){Newberry}, {Boroson}, \&
  {Kirshner}}]{newberry:90}
{Newberry}, M.~V., {Boroson}, T.~A., \& {Kirshner}, R.~P. 1990, \apj, 350, 585

\bibitem[{Oke {et~al.}(1995)Oke, Cohen, Carr, Cromer, Dingizian, Harris,
  Labrecque, Luciano, Schaal, Epps, \& Miller}]{oke:95}
Oke, J.~B., Cohen, J.~G., Carr, M., Cromer, J., Dingizian, A., Harris, F.~H.,
  Labrecque, S., Luciano, R., Schaal, W., Epps, H., \& Miller, J. 1995, \pasp,
  107, 375

\bibitem[{{Osterbrock} {et~al.}(1996){Osterbrock}, {Fulbright}, {Martel},
  {Keane}, {Trager}, \& {Basri}}]{osterbrock:96}
{Osterbrock}, D.~E., {Fulbright}, J.~P., {Martel}, A.~R., {Keane}, M.~J.,
  {Trager}, S.~C., \& {Basri}, G. 1996, \pasp, 108, 277+

\bibitem[{{Peebles}(1970)}]{peebles:70}
{Peebles}, P.~J.~E. 1970, \aj, 75, 13

\bibitem[{{Poggianti} {et~al.}(2001){Poggianti}, {Bridges}, {Carter},
  {Mobasher}, {Doi}, {Iye}, {Kashikawa}, {Komiyama}, {Okamura}, {Sekiguchi},
  {Shimasaku}, {Yagi}, \& {Yasuda}}]{poggianti:01b}
{Poggianti}, B.~M., {Bridges}, T.~J., {Carter}, D., {Mobasher}, B., {Doi}, M.,
  {Iye}, M., {Kashikawa}, N., {Komiyama}, Y., {Okamura}, S., {Sekiguchi}, M.,
  {Shimasaku}, K., {Yagi}, M., \& {Yasuda}, N. 2001, \apj, 563, 118

\bibitem[{{Poggianti} {et~al.}(1999){Poggianti}, {Smail}, {Dressler}, {Couch},
  {Barger}, {Butcher}, {Ellis}, \& {Oemler}}]{poggianti:99}
{Poggianti}, B.~M., {Smail}, I., {Dressler}, A., {Couch}, W.~J., {Barger},
  A.~J., {Butcher}, H., {Ellis}, R.~S., \& {Oemler}, A.~J. 1999, \apj, 518, 576

\bibitem[{{Postman} {et~al.}(2005){Postman}, {Franx}, {Cross}, {Holden},
  {Ford}, {Illingworth}, \& {et al.}}]{postman:05}
{Postman}, M., {Franx}, M., {Cross}, N.~J.~G., {Holden}, B., {Ford}, H.~C.,
  {Illingworth}, G.~D., \& {et al.} 2005, \apj, 623, 721

\bibitem[{{Press} {et~al.}(1992){Press}, {Teukolsky}, {Vetterling}, \&
  {Flannery}}]{press:92}
{Press}, W.~H., {Teukolsky}, S.~A., {Vetterling}, W.~T., \& {Flannery}, B.~P.
  1992, {Numerical recipes in FORTRAN. The art of scientific computing}
  (Cambridge: University Press, |c1992, 2nd ed.)

\bibitem[{{Ramella} {et~al.}(1989){Ramella}, {Geller}, \&
  {Huchra}}]{ramella:89}
{Ramella}, M., {Geller}, M.~J., \& {Huchra}, J.~P. 1989, \apj, 344, 57

\bibitem[{{Sandage} \& {Visvanathan}(1978)}]{sandage:78}
{Sandage}, A. \& {Visvanathan}, N. 1978, \apj, 225, 742

\bibitem[{{Smail} {et~al.}(2001){Smail}, {Kuntschner}, {Kodama}, {Smith},
  {Packham}, {Fruchter}, \& {Hook}}]{smail:01}
{Smail}, I., {Kuntschner}, H., {Kodama}, T., {Smith}, G.~P., {Packham}, C.,
  {Fruchter}, A.~S., \& {Hook}, R.~N. 2001, \mnras, 323, 839

\bibitem[{{Smail} {et~al.}(1999){Smail}, {Morrison}, {Gray}, {Owen}, {Ivison},
  {Kneib}, \& {Ellis}}]{smail:99}
{Smail}, I., {Morrison}, G., {Gray}, M.~E., {Owen}, F.~., {Ivison}, R.~J.,
  {Kneib}, J.-P., \& {Ellis}, R.~S. 1999, \apj, 525, 609

\bibitem[{{Stanford} {et~al.}(1998){Stanford}, {Eisenhardt}, \&
  {Dickinson}}]{stanford:98}
{Stanford}, S.~A., {Eisenhardt}, P.~R., \& {Dickinson}, M. 1998, \apj, 492, 461

\bibitem[{{Terlevich} {et~al.}(2001){Terlevich}, {Caldwell}, \&
  {Bower}}]{terlevich:01}
{Terlevich}, A.~I., {Caldwell}, N., \& {Bower}, R.~G. 2001, \mnras, 326, 1547

\bibitem[{{Terlevich} {et~al.}(1999){Terlevich}, {Kuntschner}, {Bower},
  {Caldwell}, \& {Sharples}}]{terlevich:99}
{Terlevich}, A.~I., {Kuntschner}, H., {Bower}, R.~G., {Caldwell}, N., \&
  {Sharples}, R.~M. 1999, \mnras, 310, 445

\bibitem[{{Thomas} {et~al.}(2005){Thomas}, {Maraston}, {Bender}, \& {de
  Oliveira}}]{thomas:05}
{Thomas}, D., {Maraston}, C., {Bender}, R., \& {de Oliveira}, C.~M. 2005, \apj,
  621, 673

\bibitem[{{Trager} {et~al.}(2000){Trager}, {Faber}, {Worthey}, \& {Gonz{\'
  a}lez}}]{trager:00}
{Trager}, S.~C., {Faber}, S.~M., {Worthey}, G., \& {Gonz{\' a}lez}, J.~J. 2000,
  \aj, 120, 165

\bibitem[{{Tran} {et~al.}(2003{\natexlab{a}}){Tran}, {Franx}, {Illingworth},
  {Kelson}, \& {van Dokkum}}]{tran:03b}
{Tran}, K.~H., {Franx}, M., {Illingworth}, G., {Kelson}, D.~D., \& {van
  Dokkum}, P. 2003{\natexlab{a}}, \apj, 599, 865

\bibitem[{{Tran} {et~al.}(2004){Tran}, {Franx}, {Illingworth}, {van Dokkum},
  {Kelson}, \& {Magee}}]{tran:04a}
{Tran}, K.~H., {Franx}, M., {Illingworth}, G.~D., {van Dokkum}, P., {Kelson},
  D.~D., \& {Magee}, D. 2004, \apj, 609, 683

\bibitem[{{Tran} {et~al.}(1999){Tran}, {Kelson}, {van Dokkum}, {Franx},
  {Illingworth}, \& {Magee}}]{tran:99}
{Tran}, K.~H., {Kelson}, D.~D., {van Dokkum}, P., {Franx}, M., {Illingworth},
  G.~D., \& {Magee}, D. 1999, \apj, 522

\bibitem[{{Tran} {et~al.}(2003{\natexlab{b}}){Tran}, {Simard}, {Illingworth},
  \& {Franx}}]{tran:03a}
{Tran}, K.~H., {Simard}, L., {Illingworth}, G., \& {Franx}, M.
  2003{\natexlab{b}}, \apj, 590, 238

\bibitem[{{Tran} {et~al.}(2005{\natexlab{a}}){Tran}, {van Dokkum},
  {Illingworth}, {Kelson}, {Gonzalez}, \& {Franx}}]{tran:05a}
{Tran}, K.~H., {van Dokkum}, P., {Illingworth}, G.~D., {Kelson}, D.,
  {Gonzalez}, A., \& {Franx}, M. 2005{\natexlab{a}}, \apj, 619, 134

\bibitem[{{Tran} {et~al.}(2005{\natexlab{b}}){Tran}, {van Dokkum}, {Franx},
  {Illingworth}, {Kelson}, \& {Schreiber}}]{tran:05b}
{Tran}, K.-V.~H., {van Dokkum}, P., {Franx}, M., {Illingworth}, G.~D.,
  {Kelson}, D.~D., \& {Schreiber}, N.~M.~F. 2005{\natexlab{b}}, \apjl, 627, L25

\bibitem[{Valdes(1982)}]{valdes:82}
Valdes, F. 1982, Faint Object Classification and Analysis System Manual

\bibitem[{{van Dokkum} {et~al.}(2000){van Dokkum}, {Franx}, {Fabricant},
  {Illingworth}, \& {Kelson}}]{vandokkum:00}
{van Dokkum}, P.~G., {Franx}, M., {Fabricant}, D., {Illingworth}, G.~D., \&
  {Kelson}, D.~D. 2000, \apj, 541, 95

\bibitem[{{van Dokkum} {et~al.}(1999){van Dokkum}, {Franx}, {Fabricant},
  {Kelson}, \& {Illingworth}}]{vandokkum:99}
{van Dokkum}, P.~G., {Franx}, M., {Fabricant}, D., {Kelson}, D.~D., \&
  {Illingworth}, G.~D. 1999, \apjl, 520, L95

\bibitem[{{van Dokkum} {et~al.}(1998){van Dokkum}, {Franx}, {Kelson},
  {Illingworth}, {Fisher}, \& {Fabricant}}]{vandokkum:98a}
{van Dokkum}, P.~G., {Franx}, M., {Kelson}, D.~D., {Illingworth}, G.~D.,
  {Fisher}, D., \& {Fabricant}, D. 1998, \apj, 500, 714+

\bibitem[{{Worthey} \& {Ottaviani}(1997)}]{worthey:97}
{Worthey}, G. \& {Ottaviani}, D.~L. 1997, \apjs, 111, 377

\bibitem[{{Wuyts} {et~al.}(2004){Wuyts}, {van Dokkum}, {Kelson}, {Franx}, \&
  {Illingworth}}]{wuyts:04}
{Wuyts}, S., {van Dokkum}, P.~G., {Kelson}, D.~D., {Franx}, M., \&
  {Illingworth}, G.~D. 2004, \apj, 605, 677

\bibitem[{{Yee} {et~al.}(1996){Yee}, {Ellingson}, \& {Carlberg}}]{yee:96}
{Yee}, H.~K.~C., {Ellingson}, E., \& {Carlberg}, R.~G. 1996, \apjs, 102, 269+

\end{thebibliography}

\clearpage
\pagestyle{empty}
\begin{deluxetable}{lrlrrrrrrrrrrrrr}
\rotate
\tablecolumns{15}
\tablewidth{0pc}
\tablecaption{MS 1054--03 Members\label{table1}}
\tablehead{
  \colhead{HST\#} & \colhead{z}  & 
  \colhead{$Q_z$\tablenotemark{a}} &
  \colhead{B\#\tablenotemark{b}} &
  \colhead{$i_{775}$\tablenotemark{b}} &
  \colhead{$(V-i_{775})$\tablenotemark{b}} &
  \colhead{T-type\tablenotemark{c}}  &
  \colhead{$\Delta$(RA)\tablenotemark{d}} &
  \colhead{$\Delta$(Dec)\tablenotemark{d}}&
  \colhead{OII\tablenotemark{e}} & \colhead{$\pm$} &
  \colhead{H$\delta$\tablenotemark{e}} & \colhead{$\pm$} &
  \colhead{H$\gamma$\tablenotemark{e}} & \colhead{$\pm$} &
  \colhead{$\sigma_{1D}$\tablenotemark{f}} }
\startdata
Absorption\\                               
H1192 & 0.8390 & 3 & B5324 & 20.77 & 1.675 & 0(99) & 136.6 & -57.9 & -0.2 & 0.7 & 0.2 & 0.6 & 1.5 & 0.7 & 137$^{\dag}$ \\ 
H1464 & 0.8391 & 3 & B8438 & 21.46 & 1.604 & -2 & 130.9 & -25.2 & -1.3 & 1.6 & 2.0 & 1.2 & -2.4 & 1.4 & 174 \\ 
H1609 & 0.8296 & 3 & B8222 & 21.89 & 1.643 & -4 & 120.1 & 2.6 & 3.3 & 1.9 & -4.2 & 1.7 & -2.0 & 1.8 & 188 \\ 
H1649 & 0.8373 & 3 & B8456 & 21.11 & 1.650 & -4 & 122.6 & -16.4 & 0.4 & 1.4 & 1.0 & 0.9 & 3.1 & 1.0 & 243$^{\dag}$ \\ 
H2169 & 0.8324 & 3 & B4353 & 21.69 & 1.653 & -2 & 91.0 & -63.4 & 4.9 & 1.6 & 4.2 & 1.2 & 3.7 & 1.5 & 170 \\ 
H2409 & 0.8372 & 3 & B3718 & 21.76 & 1.736 & -4 & 84.4 & -23.7 & 0.5 & 1.5 & 1.7 & 1.2 & -3.8 & 1.3 & 287$^{\dag}$ \\ 
H2525 & 0.8323 & 3 & B4254 & 23.44 & 1.591 & -5 & 75.8 & -38.5 & -0.5 & 2.8 & -6.8 & 2.7 & 3.0 & 3.2 & 71 \\ 
H2812 & 0.8274 & 3 & B4137 & 22.95 & 1.538 & -4 & 68.3 & -24.2 & 0.8 & 2.9 & 1.9 & 2.7 & -1.7 & 3.5 & 102 \\ 
H2881 & 0.8374 & 3 & B8362 & 22.20 & 1.511 & -1 & 68.3 & 56.0 & 3.0 & 2.7 & 4.8 & 1.7 & 3.1 & 1.8 & 156 \\ 
H2977 & 0.8316 & 3 & B4778 & 22.77 & 1.518 & -2 & 61.5 & -49.4 & 2.3 & 2.1 & -4.9 & 2.1 & 5.7 & 2.4 & 110 \\ 
H3058 & 0.8309 & 3 & B4167 & 20.31 & 1.593 & -5 & 52.8 & -9.5 & 1.8 & 1.1 & 1.7 & 0.9 & 0.5 & 1.0 & 303$^{\dag}$\\  
H3105 & 0.8275 & 3 & B4324 & 21.84 & 1.539 & -5 & 64.8 & -29.3 & 0.7 & 1.0 & 3.9 & 0.9 & 1.0 & 1.1 & 142 \\ 
H3212 & 0.8392 & 3 & B2175 & 22.23 & 1.611 & -1(99) & 54.5 & -132.2 & 0.9 & 2.3 & -2.9 & 2.0 & -3.6 & 2.4 & 155 \\ 
H3288 & 0.8324 & 3 & B4363 & 23.58 & 1.680 & -5 & 49.6 & -12.1 & 7.1 & 3.6 & 2.0 & 3.0 & -4.8 & 3.8 & 125 \\ 
H3398 & 0.8343 & 3 & B4212 & 23.18 & 1.607 & -5 & 47.7 & -2.5 & -3.9 & 4.5 & -4.4 & 3.8 & 0.4 & 4.2 & 116 \\ 
H3495 & 0.8199 & 3 & B4213 & 22.07 & 1.613 & -2(99) & 46.1 & -1.5 & 1.2 & 2.2 & 3.6 & 2.5 & 2.8 & 3.1 & 190 \\ 
H3524 & 0.8195 & 3 & B4214 & 22.08 & 1.522 & -1(99) & 44.9 & -1.3 & 1.8 & 2.0 & 6.6 & 2.3 & -6.1 & 3.1 & 148 \\ 
H3532 & 0.8302 & 3 & B2787 & 22.77 & 1.634 & -4 & 42.9 & -76.5 & -2.9 & 3.2 & 0.9 & 2.7 & -11.6 & 4.0 & 110 \\ 
H3581 & 0.8352 & 3 & B4358 & 21.68 & 1.603 & -5 & 43.8 & -5.4 & 2.0 & 1.7 & 2.6 & 1.3 & 1.6 & 1.5 & 193 \\ 
H3596 & 0.8173 & 3 & B4942 & 21.82 & 1.520 & -4 & 41.3 & -34.2 & 0.6 & 0.4 & 1.7 & 0.5 & 1.5 & 0.8 & 125 \\ 
H3768 & 0.8177 & 3 & B4364 & 21.38 & 1.612 & -5 & 38.9 & -0.2 & -2.5 & 1.4 & 1.9 & 1.2 & 1.3 & 1.5 & 221$^{\dag}$\\  
H3816 & 0.8238 & 3 & B2874 & 22.15 & 1.644 & -2 & 33.5 & -56.8 & 1.2 & 1.6 & 4.1 & 1.4 & 2.6 & 1.9 & 179 \\ 
H3835 & 0.8396 & 3 & B8262 & 21.46 & 1.586 & -5 & 39.7 & 96.5 & 2.2 & 1.3 & 3.0 & 1.1 & 3.7 & 1.2 & 169 \\ 
H3995 & 0.8339 & 3 & B2390 & 21.23 & 1.633 & -5 & 30.9 & -80.9 & -0.9 & 1.3 & 1.4 & 1.1 & -0.6 & 1.3 & 229 \\ 
H4012 & 0.8309 & 3 & B2391 & 21.82 & 1.592 & -5 & 31.4 & -82.7 & -2.7 & 1.3 & 4.1 & 0.9 & -1.5 & 1.2 & 133 \\ 
H4031 & 0.8218 & 3 & B3185 & 22.14 & 1.621 & -5 & 26.1 & -31.1 & 4.1 & 1.7 & -4.9 & 2.0 & -0.5 & 2.3 & 161 \\ 
H4039 & 0.8416 & 3 & B5134 & 23.22 & 1.211 & -4 & 22.8 & -22.1 & 0.3 & 3.3 & 1.8 & 2.9 & 0.8 & 3.3 & 67 \\ 
H4222 & 0.8443 & 3 & B4620 & 21.42 & 0.970 & 6 & 22.9 & 5.9 & 2.0 & 2.2 & -9.0 & 2.4 & -2.5 & 2.9 & \nodata \\ 
H4343 & 0.8378 & 3 & B7835 & 22.13 & 1.600 & -2 & 18.3 & 161.2 & 1.0 & 2.5 & -0.6 & 1.7 & 2.6 & 1.9 & 119 \\ 
H4345 & 0.8349 & 3 & B5006 & 21.01 & 1.674 & -5 & 21.6 & -13.1 & 0.9 & 0.9 & -1.3 & 0.8 & 2.2 & 0.8 & 335$^{\dag}$\\  
H4428 & 0.8253 & 3 & B5062 & 22.18 & 1.579 & -2 & 12.1 & -5.5 & 3.9 & 1.6 & 0.4 & 1.8 & 4.6 & 2.0 & 135 \\ 
H4459/4528 & 0.8260 & 3 & B4169 & 22.93 & 1.544 & -5(99) & 10.3 & 44.6 & -0.1 & 2.1 & 2.3 & 2.0 & -0.3 & 2.2 & 93 \\ 
H4459 & 0.8265 & 3 & B4169 & 22.93 & 1.544 & -5(99) & 10.3 & 44.6 & 1.7 & 1.2 & 2.3 & 1.3 & 1.0 & 1.6 & 93 \\ 
H4520 & 0.8316 & 3 & B2830 & 19.87 & 1.705 & -5 & 0.0 & 0.0 & 0.2 & 1.0 & 0.7 & 0.7 & 1.0 & 0.7 & 322$^{\dag}$\\  
H4715 & 0.8339 & 3 & B5099 & 22.49 & 1.639 & -2 & 2.1 & 4.3 & 0.3 & 3.1 & 5.2 & 2.7 & -2.1 & 2.7 & 172 \\ 
H4822 & 0.8333 & 3 & B3148 & 22.43 & 1.494 & 10 & -0.4 & 3.2 & 5.0 & 3.2 & -2.6 & 2.7 & 3.7 & 2.8 & 50 \\ 
H4870 & 0.8308 & 3 & B2589 & 21.81 & 1.544 & -5 & -3.6 & -31.3 & 1.8 & 1.7 & 1.9 & 1.5 & 1.6 & 1.8 & 281$^{\dag}$\\  
H4922 & 0.8280 & 3 & B5116 & 22.32 & 1.644 & -4 & -5.7 & 12.5 & -1.0 & 2.2 & -1.6 & 2.0 & 3.9 & 2.2 & 168 \\ 
H4926 & 0.8252 & 3 & B3035 & 21.77 & 1.719 & -3 & -4.2 & -3.7 & 1.0 & 1.6 & 0.8 & 1.4 & -0.6 & 1.6 & 310$^{\dag}$\\  
H5111 & 0.8186 & 3 & B2301 & 22.20 & 1.558 & -5 & -13.8 & -37.4 & 2.0 & 1.7 & 3.5 & 1.6 & 1.7 & 2.0 & 131 \\ 
H5280 & 0.8301 & 3 & B3070 & 21.63 & 1.649 & -2 & -21.6 & 20.6 & 1.0 & 1.9 & -1.1 & 1.5 & 3.4 & 1.8 & 259$^{\dag}$\\  
H5196 & 0.8418 & 3 & B2741 & 21.39 & 1.645 & -2 & -16.6 & 0.2 & 2.4 & 1.5 & 5.8 & 1.2 & 0.5 & 1.5 & 301$^{\dag}$\\  
H5216 & 0.8351 & 3 & B2403 & 21.31 & 1.555 & 1 & -24.1 & -5.6 & -3.3 & 3.2 & 3.9 & 1.7 & 3.4 & 2.0 & 152 \\ 
H5226 & 0.8348 & 3 & B5072 & 22.17 & 1.618 & -2 & -18.8 & 31.1 & -2.5 & 2.9 & 1.9 & 1.7 & 0.5 & 2.2 & 204 \\ 
H5269 & 0.8243 & 3 & B2901 & 22.28 & 1.551 & -5 & -21.7 & 9.3 & 3.0 & 1.6 & -3.3 & 1.7 & -0.6 & 2.0 & 163 \\ 
H5298 & 0.8311 & 3 & B3031 & 21.95 & 1.593 & -2 & -21.9 & 19.3 & 1.3 & 2.1 & 1.2 & 1.7 & 0.9 & 2.1 & 283$^{\dag}$ \\ 
H5325 & 0.8314 & 3 & B1509 & 22.37 & 1.592 & -5 & -24.5 & -79.5 & -0.9 & 4.2 & 0.4 & 2.8 & -6.8 & 3.5 & 102 \\ 
H5338 & 0.8293 & 3 & B3131 & 22.58 & 1.630 & -4 & -24.3 & 23.7 & -1.5 & 2.0 & 2.2 & 1.7 & 2.1 & 2.0 & 113 \\ 
H5347 & 0.8251 & 3 & B4987 & 21.02 & 1.544 & -1(99) & -34.6 & 56.3 & 0.5 & 1.9 & 3.1 & 1.8 & 3.3 & 2.3 & 253$^{\dag}$ \\ 
H5432 & 0.8149 & 3 & B2651 & 22.39 & 1.252 & -1 & -29.6 & 2.3 & -1.8 & 0.8 & 10.4 & 0.6 & -8.8 & 1.0 & 62 \\ 
H5442 & 0.8201 & 3 & B2639 & 22.35 & 1.549 & -5 & -29.9 & 4.5 & -0.4 & 2.5 & 0.1 & 2.3 & 5.2 & 2.7 & 121 \\ 
H5450 & 0.8365 & 3 & B2143 & 20.58 & 1.612 & -5 & -46.4 & -3.6 & -1.0 & 2.0 & 1.0 & 1.1 & 0.1 & 1.2 & 233$^{\dag}$\\  
H5529 & 0.8218 & 3 & B2894 & 21.56 & 1.643 & -5 & -38.0 & 31.2 & -2.2 & 1.3 & 2.4 & 1.1 & 2.5 & 1.4 & 182$^{\dag}$\\  
H5541 & 0.8276 & 3 & B3241 & 22.57 & 1.610 & -1 & -36.1 & 45.8 & -4.2 & 3.2 & 2.1 & 2.6 & 4.8 & 2.6 & 132 \\ 
H5543 & 0.8246 & 3 & B4988 & 21.96 & 1.604 & -5 & -35.4 & 55.5 & -0.4 & 3.0 & 2.7 & 2.8 & -1.5 & 3.5 & 220 \\ 
H5577 & 0.8310 & 3 & B3052 & 21.54 & 1.597 & -4 & -43.8 & 48.1 & 2.5 & 1.4 & 0.0 & 1.1 & 4.5 & 1.3 & 305$^{\dag}$\\  
H5607 & 0.8294 & 3 & B3113 & 22.19 & 1.649 & -5 & -42.1 & 46.0 & 4.4 & 4.5 & 5.3 & 4.0 & -3.3 & 5.2 & 138 \\ 
H5666 & 0.8315 & 3 & B830 & 20.82 & 1.641 & -1 & -57.8 & -83.9 & 2.0 & 1.2 & 1.3 & 0.9 & 0.4 & 1.0 & 285$^{\dag}$\\  
H5672 & 0.8306 & 3 & B2709 & 21.88 & 1.597 & -5 & -46.8 & 30.2 & 0.6 & 3.8 & 8.4 & 2.3 & -10.6 & 3.8 & 180 \\ 
H5720 & 0.8240 & 3 & B4374 & 22.01 & 1.538 & -4 & -44.0 & 100.9 & -1.4 & 1.6 & 5.3 & 1.4 & -1.3 & 1.8 & 178 \\ 
H5756 & 0.8305 & 3 & B4766 & 21.22 & 1.628 & -5 & -59.7 & 97.7 & -2.4 & 1.2 & 1.4 & 0.9 & -1.7 & 1.0 & 231$^{\dag}$ \\ 
H5757 & 0.8207 & 3 & B2257 & 21.65 & 1.616 & -5 & -49.6 & 1.3 & -3.3 & 1.6 & -0.2 & 1.3 & 1.9 & 1.5 & 167 \\ 
H6036/6064 & 0.8333 & 3 & B1496 & 21.41 & 1.619 & -5(99) & -63.5 & -29.3 & -0.3 & 1.4 & 3.4 & 0.9 & 1.3 & 1.0 & 253$^{\dag}$ \\ 
H6038 & 0.8312 & 3 & B2353 & 21.17 & 1.641 & -5 & -61.7 & 28.1 & -0.2 & 0.4 & 1.2 & 0.3 & -0.2 & 0.4 & 253 \\ 
H6048 & 0.8304 & 3 & B2041 & 23.25 & 1.463 & -5 & -63.1 & -0.5 & 1.3 & 4.0 & 0.5 & 3.1 & -0.3 & 3.8 & 136 \\ 
H6064 & 0.8341 & 3 & B1461 & 21.56 & 1.541 & -5(99) & -63.6 & -29.8 & 1.7 & 0.9 & 3.4 & 0.8 & 0.6 & 0.8 & 181 \\ 
H6301 & 0.8277 & 3 & \nodata & \nodata & \nodata & \nodata & -70.8 & 25.2 & 1.6 & 1.3 & 3.4 & 1.1 & 0.2 & 1.2 & 248$^{\dag}$ \\ 
H6412 & 0.8322 & 3 & B2385 & 22.75 & 1.702 & -1 & -70.8 & 35.5 & 2.1 & 3.9 & 4.4 & 3.0 & -2.0 & 4.0 & 102 \\ 
H6516 & 0.8289 & 3 & B1644 & 22.98 & 0.913 & -2 & -75.8 & -9.4 & 0.3 & 3.0 & 0.2 & 3.3 & -2.9 & 4.1 & 37 \\ 
H6567 & 0.8265 & 3 & B654 & 21.46 & 1.614 & 1(99) & -89.2 & -59.9 & -4.6 & 3.2 & 4.3 & 2.0 & 2.4 & 2.4 & 201$^{\dag}$\\  
H6645 & 0.8213 & 3 & B1111 & 22.23 & 1.166 & -2 & -85.1 & -34.0 & -3.4 & 1.8 & 6.4 & 2.0 & 1.3 & 2.8 & 64 \\ 
H7079 & 0.8351 & 3 & B984 & 21.86 & 1.564 & -5 & -118.8 & -0.8 & 0.1 & 0.8 & 3.0 & 0.6 & 1.7 & 0.8 & 132 \\ 
H7746 & 0.8252 & 3 & B2315 & 21.68 & 1.702 & -2 & -134.2 & 111.0 & 0.3 & 1.3 & -1.4 & 1.2 & -1.5 & 1.3 & 268 \\ 
Post-starburst\\                               
H2710 & 0.8420 & 3 & B3224 & 21.85 & 1.588 & -4 & 75.8 & -88.8 & -2.7 & 2.3 & 3.3 & 1.7 & 7.1 & 2.0 & 140 \\ 
H2872 & 0.8340 & 3 & B4488 & 21.58 & 1.454 & 0 & 67.6 & -40.7 & -1.3 & 1.4 & 6.1 & 1.1 & 3.2 & 1.4 & 179$^{\dag}$\\  
H3910 & 0.8332 & 3 & B4735 & 21.41 & 1.975 & -2 & 32.1 & -11.7 & -0.5 & 1.3 & 2.6 & 1.0 & 6.2 & 1.3 & 295$^{\dag}$\\  
H4389 & 0.8292 & 3 & B4366 & 21.78 & 1.131 & 8(99) & 14.3 & 29.8 & -3.9 & 3.1 & 8.5 & 2.2 & 5.9 & 2.9 & \nodata \\ 
H4390 & 0.8240 & 3 & B4305 & 22.55 & 1.573 & -5 & 13.3 & 34.1 & 2.8 & 2.5 & 4.6 & 2.2 & 6.0 & 2.8 & \nodata \\ 
H5359 & 0.8238 & 3 & B2649 & 20.63 & 1.460 & 1 & -26.4 & 6.9 & 0.4 & 1.0 & 5.6 & 0.8 & 6.0 & 1.0 & 138$^{\dag}$ \\ 
H5534 & 0.8395 & 3 & B2937 & 21.33 & 1.519 & -4 & -38.2 & 34.1 & -1.4 & 1.3 & 5.6 & 1.0 & 3.4 & 1.2 & 152 \\ 
H5691 & 0.8331 & 3 & B2476 & 22.64 & 1.643 & -1 & -44.3 & 11.5 & -0.5 & 3.2 & 3.0 & 2.4 & 8.5 & 2.2 & 152 \\ 
H5786 & 0.8344 & 3 & B2178 & 22.38 & 1.604 & -2 & -51.0 & -4.4 & 10.9 & 3.8 & 1.6 & 2.3 & 9.7 & 2.5 & 140 \\ 
H5833 & 0.8262 & 3 & B1859 & 22.24 & 1.530 & 1 & -54.9 & -19.0 & 0.5 & 1.3 & 0.3 & 1.3 & 7.8 & 1.7 & 86 \\ 
H5840 & 0.8259 & 3 & B2468 & 21.05 & 1.487 & -1(99) & -52.4 & 22.9 & -1.3 & 0.6 & 7.4 & 0.5 & 7.7 & 0.6 & 211$^{\dag}$\\  
H5923 & 0.8392 & 3 & B3180 & 21.25 & 1.513 & 3 & -57.8 & 73.2 & -3.7 & 1.4 & 8.1 & 0.9 & 6.6 & 1.2 & 165 \\ 
H5926 & 0.8231 & 3 & B2333 & 22.40 & 1.386 & -1 & -55.2 & 16.6 & -2.4 & 2.8 & 11.8 & 2.2 & 10.0 & 3.0 & 68 \\ 
H6164 & 0.8257 & 3 & B2328 & 22.01 & 1.643 & -5 & -63.5 & 28.1 & 0.2 & 5.1 & 4.2 & 4.4 & 4.5 & 5.4 & 174 \\ 
H6309 & 0.8284 & 3 & B1772 & 22.77 & 1.430 & -4 & -67.3 & -10.0 & 1.4 & 2.9 & 6.5 & 2.6 & 5.4 & 3.5 & 56 \\ 
H8001 & 0.8444 & 3 & B2226 & 25.52 & 1.144 & 9 & -111.9 & 69.5 & -2.7 & 2.1 & 10.3 & 1.3 & 9.6 & 1.7 & 165 \\ 
H987 & 0.8251 & 3 & B6304 & 21.74 & 1.712 & -2 & 152.4 & 81.4 & -2.4 & 1.1 & 4.7 & 0.9 & 4.1 & 1.1 & 221 \\ 
Emission\\                               
H1242 & 0.8286 & 3 & B3386 & 22.67 & 0.830 & 4 & 137.3 & -70.9 & -40.6 & 3.3 & 8.3 & 1.9 & -14.4 & 3.0 & 29 \\ 
H1532 & 0.8219 & 3 & B6995 & 21.82 & 0.875 & 6(99) & 128.9 & 86.4 & -9.4 & 3.6 & 1.0 & 3.0 & 11.6 & 4.2 & 35 \\ 
H159 & 0.8228 & 3 & B6011 & 22.57 & 0.857 & 6 & 208.7 & 50.1 & -24.4 & 2.3 & 5.5 & 2.0 & -3.5 & 2.9 & 21 \\ 
H2334 & 0.8453 & 3 & B4015 & 22.33 & 1.473 & -1 & 82.3 & -35.4 & -5.9 & 2.6 & 4.6 & 1.5 & 1.2 & 1.9 & 120 \\ 
H2377 & 0.8369 & 3 & B8422 & 21.86 & 1.359 & 1 & 85.1 & 31.5 & -6.9 & 2.2 & 6.5 & 1.5 & 6.6 & 2.0 & 79 \\ 
H2538 & 0.8179 & 3 & B3739 & 23.16 & 1.493 & -5 & 75.3 & -12.8 & -12.0 & 3.3 & -1.0 & 2.1 & -14.1 & 3.0 & 106 \\ 
H255 & 0.8346 & 3 & B6736 & 22.22 & 1.041 & 3 & 187.8 & 35.0 & -23.1 & 5.8 & 4.1 & 3.2 & 2.2 & 5.8 & 24 \\ 
H2609 & 0.8362 & 3 & B3686 & 22.67 & 1.587 & -4 & 75.2 & -11.0 & -5.6 & 4.5 & 3.8 & 2.6 & 6.5 & 3.4 & \nodata \\ 
H3043 & 0.8278 & 3 & B4308 & 23.48 & 1.403 & -4 & 59.9 & -22.5 & -6.6 & 4.3 & 4.4 & 3.4 & 11.3 & 4.5 & 64 \\ 
H3284 & 0.8449 & 3 & B2503 & 21.41 & 1.005 & 4 & 53.9 & -93.9 & -5.3 & 1.7 & 3.9 & 1.5 & -1.8 & 1.9 & 48 \\ 
H3585 & 0.8366 & 3 & B3795 & 23.83 & 1.426 & -1 & 39.8 & 27.1 & -16.1 & 9.6 & 7.1 & 5.9 & -11.7 & 8.3 & 37 \\ 
H4223 & 0.8303 & 3 & B2844 & 22.48 & 0.905 & -1(99) & 18.4 & -41.1 & -5.1 & 2.6 & 7.5 & 2.1 & 2.5 & 2.8 & 34 \\ 
H4683/4741 & 0.8329 & 3 & B3812 & 21.65 & 0.274 & 8(99) & 6.9 & 65.5 & -56.8 & 1.5 & -1.2 & 0.9 & -6.2 & 1.2 & \nodata \\ 
H4683 & 0.8332 & 3 & B3812 & 21.65 & 0.274 & 8(99) & 6.9 & 65.5 & -60.7 & 2.4 & -3.8 & 1.5 & \nodata & 0.0 & \nodata \\ 
H4705 & 0.8453 & 3 & B4954 & 21.22 & 1.289 & 1 & 6.0 & 8.4 & -5.6 & 1.2 & 6.1 & 0.9 & 5.7 & 1.0 & 253$^{\dag}$\\  
H5599 & 0.8112 & 3 & B1978 & 21.95 & 0.924 & 4 & -57.0 & -5.6 & -13.7 & 2.4 & 2.4 & 1.6 & 0.7 & 2.1 & \nodata \\ 
H5955 & 0.8293 & 3 & B1814 & 22.65 & 1.423 & -5 & -60.5 & -14.0 & -5.3 & 4.3 & 5.3 & 3.2 & 12.9 & 4.4 & \nodata \\ 
H5992 & 0.8306 & 3 & B1679 & 22.44 & 1.125 & 1 & -58.1 & -25.2 & -5.3 & 3.1 & 0.5 & 2.5 & -6.5 & 4.0 & 39 \\ 
H6372 & 0.8332 & 3 & B754 & 22.79 & 1.324 & 3 & -75.5 & -73.5 & -8.0 & 6.4 & 6.3 & 2.7 & 7.9 & 3.4 & 42 \\ 
H6695 & 0.8352 & 3 & B845 & 22.50 & 1.743 & -2 & -90.2 & -44.3 & -7.1 & 3.7 & -1.8 & 2.2 & -48.1 & 3.3 & 150 \\ 
H6812 & 0.8397 & 3 & B2218 & 22.76 & 0.922 & 3 & -97.9 & 56.7 & -21.0 & 3.8 & 3.1 & 3.6 & 2.3 & 4.6 & \nodata \\ 
H7107 & 0.8412 & 3 & B2991 & 22.45 & 0.719 & 1 & -106.2 & 116.8 & -40.4 & 3.3 & -4.2 & 2.7 & 10.3 & 2.8 & \nodata \\ 
H7163 & 0.8255 & 3 & B156 & 23.54 & 0.843 & 0 & -186.0 & 12.4 & -34.7 & 8.0 & 8.6 & 5.5 & 5.1 & 6.6 & 17 \\ 
H7186 & 0.8246 & 3 & B24 & 22.80 & 0.811 & -1 & -203.5 & 24.1 & -26.3 & 3.3 & 5.1 & 2.7 & 8.4 & 4.0 & 18 \\ 
H7333 & 0.8446 & 3 & B2926 & 24.04 & 0.254 & 8 & -119.9 & 127.8 & -42.7 & 10.0 & 18.2 & 11.9 & \nodata & \nodata & 9 \\ 
H7335 & 0.8280 & 3 & B215 & 22.16 & 0.806 & 4 & -174.8 & 5.6 & -33.5 & 5.3 & 10.5 & 3.8 & 2.5 & 5.9 & 32 \\ 
No Measured [OII]\\                              
H1942 & 0.8310 & 3 & B3652 & 22.01 & 1.578 & -1 & 99.2 & -38.5 & \nodata & \nodata & \nodata & \nodata & -1.1 & 1.0 & 164 \\ 
H2203 & 0.8314 & 3 & B4049 & 22.80 & 1.574 & -2 & 85.9 & -41.0 & \nodata & \nodata & 0.4 & 2.4 & -4.0 & 3.1 & 116 \\ 
H6688 & 0.8353 & 3 & B846 & 21.32 & 1.628 & -5 & -91.6 & -42.0 & \nodata & \nodata & 1.3 & 1.0 & 1.7 & 1.2 & 274$^{\dag}$ \\ 
H6690 & 0.8241 & 3 & B3003 & 22.50 & 1.366 & 4 & -85.4 & 91.7 & \nodata & \nodata & -0.5 & 1.8 & 2.5 & 2.3 & \nodata \\ 
H7228 & 0.8049 & 3 & B308 & 23.77 & 1.425 & -5 & -108.9 & -70.0 & \nodata & \nodata & 4.4 & 9.4 & 1.8 & 9.1 & 65 \\ 
Probable Members \\
H7901  & 0.8310   & 2   & B311   & 22.93  & 0.686  & 3      & -175.7 & 11.3   & -4.4   & 2.8    & 3.4    & 2.9    & 12.9   & 4.0   & \nodata \\ 
H7210  & 0.8270   & 2   & B732   & 23.24  & 0.880  & 3      & -184.4 & 54.6   & -31.3  & 9.6    & 14.6   & 6.4    & 6.7    & 8.4   & \nodata \\ 
H7076  & 0.8456   & 2   & B2225  & 23.44  & 0.626  & -1     & -104.5 & 65.7   & -0.8   & 7.1    & 24.6   & 8.9    & 10.4   & 12.1  & \nodata \\ 
H6829  & 0.8294   & 2   & B2898  & 23.05  & 0.740  & 1      & -91.2  & 92.0   & -43.2  & 6.6    & 1.2    & 4.9    & 5.1    & 5.5   & \nodata \\ 
H6065  & 0.8253   & 2   & B3692  & 22.21  & 1.167  & 6      & -58.2  & 150.4  & -11.2  & 3.5    & 5.9    & 2.9    & 0.4    & 3.4   & \nodata \\ 
H5922  & 0.8318   & 2   & B4717  & 23.40  & 1.358  & 8      & -60.2  & 102.5  & -7.3   & 4.6    & 8.0    & 4.0    & -1.1   & 7.4   & \nodata \\ 
H5555  & 0.8209   & 2   & B2577  & 22.45  & 1.551  & -4     & -38.0  & 8.1    & 0.6    & 2.2    & 4.0    & 2.3    & -4.2   & 2.8   & \nodata \\ 
H4165  & 0.8315   & 2   & B5048  & 23.61  & 1.153  & -4     & 18.2   & -11.2  & 5.2    & 4.7    & 5.9    & 4.9    & 17.1   & 6.4   & \nodata \\ 
H3447  & 0.8381   & 2   & B2873  & 23.12  & 1.445  & 3      & 45.6   & -74.7  & -3.8   & 3.6    & -5.6   & 2.7    & -11.5  & 2.9   & \nodata \\ 
H3356  & 0.8358   & 2   & B2896  & 22.97  & 1.446  & -4     & 49.0   & -79.0  & 3.6    & 3.3    & 1.7    & 3.0    & 10.0   & 3.8   & \nodata \\ 
H2746  & 0.8307   & 2   & B4063  & 23.09  & 1.468  & -2     & 70.0   & -22.9  & 1.2    & 10.6   & 24.6   & 10.2   & -6.2   & 14.6  & \nodata \\ 
H2467  & 0.8433   & 2   & B5472  & 23.37  & 1.620  & -4     & 77.7   & 24.1   & 5.0    & 5.9    & 12.9   & 4.2    & -2.4   & 6.5   & \nodata \\ 
H1724  & 0.8374   & 2   & B8312  & 22.72  & 1.635  & -1     & 111.0  & 9.4    & -11.4  & 7.4    & 7.1    & 4.3    & 5.7    & 5.1   & \nodata \\ 
H925   & 0.8245   & 2   & B5532  & 22.88  & 1.662  & -1     & 151.9  & -62.0  & -1.9   & 2.5    & -6.5   & 2.2    & 9.6    & 2.3   & \nodata \\
H426   & 0.8263   & 2   & B7431  & 22.78  & 1.279  & -1     & 186.6  & -14.5  & -4.7   & 4.3    & 6.3    & 4.1    & 8.6    & 5.5   & \nodata \\
H7603  & 0.8288   & 1   & B1192  & 23.55  & 0.708  & 3      & -140.6 & 39.4   & -57.5  & 19.1   & 5.7    & 11.0   & \nodata & \nodata & \nodata \\ 
H7212  & 0.8283   & 1   & B632   & 23.10  & 0.662  & 4      & -180.3 & 47.3   & -40.7  & 5.2    & 8.5    & 3.7    & -1.5   & 5.6   & \nodata \\ 
H6993  & 0.8291   & 1   & B2484  & 23.74  & 0.688  & 3      & -100.5 & 78.1   & \nodata & \nodata & 14.3   & 11.5   & 19.1   & 9.1  & \nodata  \\ 
H6663  & 0.8372   & 1   & B2240  & 23.67  & 0.789  & 0      & -84.4  & 41.5   & -32.0  & 9.0    & -15.3  & 12.8   & 9.8    & 13.2  & \nodata \\ 
H5990  & 0.8334   & 1   & B5082  & 22.70  & 0.442  & 8      & -67.5  & 89.5   & -8.6   & 3.3    & -1.0   & 4.7    & 0.3    & 5.9   & \nodata \\ 
H5935  & 0.8369   & 1   & B2547  & 22.60  & 1.026  & 3      & -53.6  & 27.5   & -0.7   & 2.9    & 5.9    & 2.7    & 2.3    & 3.2   & \nodata \\ 
H5545  & 0.8229   & 1   & B2356  & 23.66  & 0.499  & 8      & -36.6  & -5.9   & \nodata & \nodata & \nodata & \nodata & 45.8   & 14.9  & \nodata \\ 
H4028  & 0.8286   & 1   & B5447  & 22.46  & 0.847  & 3      & 29.6   & 83.3   & -80.7  & 23.9   & -5.4   & 8.9    & -13.9  & 8.4   & \nodata \\ 
H1266  & 0.8264   & 1   & B8107  & 23.37  & 0.665  & -1     & 135.8  & -3.2   & -32.2  & 5.4    & -4.8   & 4.2    & -0.6   & 4.3  & \nodata  \\ 
\enddata
\tablenotetext{a}{In our analysis, we consider only members with a redshift 
quality flag of 3.  The typical redshift errors from the cross-correlation
are $\sim30$ km~s$^{-1}$, and we find that the median error estimated from
galaxies observed in multiple masks is 50\kms.  
Candidate members with $Q_z=2$ or 1 are listed at the end of the table.}
\tablenotetext{b}{The identification number and corresponding ACS
photometry for the cluster galaxies are from \citet{blakeslee:06}; magnitudes
and colors (AB system) are measured within an effective radius determined
from fitting 2-D S\'ersic profiles.  At $z=0.83$,
$i_{775}$ and $(V-i_{775})$ closely correspond to rest-frame $B$ and $(U-B)$.}
\tablenotetext{c}{The morphological types as assigned by \citet{postman:05} 
using the ACS imaging; members that are also visually classified to be part of 
a merging system \citep{vandokkum:00} have ``(99)''.}
\tablenotetext{d}{The relative offset in arcseconds determined with respect to 
the brightest cluster galaxy (H4520); the BCG's J2000 coordinates are 
($10^h57^m00^s.0,$ $-3^{\circ}37'36".2$). Positive $\Delta$(RA) is west and positive 
$\Delta$(Dec) is north.}
\tablenotetext{e}{Equivalent widths and their errors are in \AA; the
indices are determined using the same bandpasses as \citet{fisher:98}.}
\tablenotetext{f}{The 27 members where the internal velocity dispersion
(km~s$^{-1}$) is measured directly \citep{wuyts:04} are daggered and have
errors of $\sim10-15$\%.  For the remaining cluster galaxies, we estimate
$\sigma_{1D}$ using the method and measurements described in \citet{tran:03b}; 
the errors on the estimated $\sigma_{1D}$, especially at
$L<L^{\ast}$, can easily be $>20$\%.}
\end{deluxetable}

\pagestyle{plaintop}
\clearpage

\begin{deluxetable}{lrrrrrrr}
\tabletypesize{\footnotesize}
\tablecolumns{7}
\tablewidth{0pc}
\tablecaption{Observed $(V-i_{775})$:  Mean Color Offset and Scatter\label{colortab}}
\tablehead{
  \colhead{Galaxy Type\tablenotemark{a}} & \colhead{$N_g$}  & 
  \colhead{Offset\tablenotemark{b}} & 
  \colhead{$\sigma_{obs}$\tablenotemark{b}} &
  \colhead{$\sigma_{int}$\tablenotemark{c}} &
  \colhead{$<V-i_{775}>$\tablenotemark{b}} & 
  \colhead{$<i_{775}-z>$\tablenotemark{b}}  }
\startdata
Absorption         & 72 & 0.001 $\pm$ 0.007 & 0.056 $\pm$ 0.006 & 0.048 $\pm$ 0.009 & 1.605 $\pm$ 0.007 & 0.675 $\pm$ 0.008 \\
~~~Absorption:  Red, Bright\tablenotemark{d} & 37 & 0.000 $\pm$ 0.008 & 0.050 $\pm$ 0.007 & 0.047 $\pm$ 0.008 & 1.616 $\pm$ 0.009 & 0.694 $\pm$ 0.013 \\
~~~Absorption:  Red, Faint\tablenotemark{d}& 30 & -0.003 $\pm$ 0.012 & 0.060 $\pm$ 0.008 & 0.055 $\pm$ 0.009 & 1.587 $\pm$ 0.013 & 0.664 $\pm$ 0.010 \\
E+A                & 17 & -0.082 $\pm$ 0.040 & 0.168 $\pm$ 0.050 & 0.152 $\pm$ 0.052 & 1.528 $\pm$ 0.038 & 0.531 $\pm$ 0.045 \\
Emission           & 26 & -0.535 $\pm$ 0.103 & 0.417 $\pm$ 0.063 & 0.407 $\pm$ 0.064 & 1.054 $\pm$ 0.104 & 0.351 $\pm$ 0.048 \\
\\
E+S0               & 94 & -0.009 $\pm$ 0.009 & 0.078 $\pm$ 0.009 & 0.072 $\pm$ 0.010 & 1.595 $\pm$ 0.009 & 0.673 $\pm$ 0.009 \\
Elliptical Only    & 55 & -0.011 $\pm$ 0.009 & 0.063 $\pm$ 0.008 & 0.055 $\pm$ 0.008 & 1.592 $\pm$ 0.011 & 0.675 $\pm$ 0.009 \\
~~~Elliptical:  Red, Bright\tablenotemark{d} & 29 & -0.008 $\pm$ 0.012 & 0.049 $\pm$ 0.010 & 0.045 $\pm$ 0.011 & 1.607 $\pm$ 0.011 & 0.697 $\pm$ 0.016 \\
~~~Elliptical:  Red, Faint\tablenotemark{d} & 25 & -0.025 $\pm$ 0.019 & 0.077 $\pm$ 0.011 & 0.071 $\pm$ 0.013 & 1.560 $\pm$ 0.026 & 0.651 $\pm$ 0.013 \\
S0 Only            & 39 & -0.005 $\pm$ 0.022 & 0.112 $\pm$ 0.030 & 0.102 $\pm$ 0.032 & 1.595 $\pm$ 0.025 & 0.659 $\pm$ 0.049 \\
~~~S0: Red, Bright\tablenotemark{d} & 13 & 0.016 $\pm$ 0.030 & 0.091 $\pm$ 0.042 & 0.085 $\pm$ 0.042 & 1.623 $\pm$ 0.037 & 0.700 $\pm$ 0.032 \\
~~~S0: Red, Faint\tablenotemark{d}  & 19 & -0.002 $\pm$ 0.024 & 0.075 $\pm$ 0.017 & 0.068 $\pm$ 0.018 & 1.596 $\pm$ 0.024 & 0.671 $\pm$ 0.028 \\
Spiral+Irr         & 26 & -0.501 $\pm$ 0.089 & 0.395 $\pm$ 0.064 & 0.273 $\pm$ 0.065 & 1.096 $\pm$ 0.085 & 0.364 $\pm$ 0.036 \\
\enddata
\tablenotetext{a}{Spectral types were assigned using [OII]$\lambda3727$, 
H$\delta$, and H$\gamma$ (see \S\ref{spectypes}).  Morphological types were 
classified using HST/ACS imaging \citep[see \S\ref{acs};][]{postman:05}.}
\tablenotetext{b}{The mean offset and observed scatter in $(V-i_{775})$ 
as well as the mean colors were determined using the biweight estimator \citep{beers:90}.}
\tablenotetext{c}{Following B06, we correct the observed scatter in color 
for measurement errors.}
\tablenotetext{d}{Galaxies are red if $\Delta(V-i_{775})\geq-0.2$.  We
divide the members into bright and faint samples using $i_{775}=22$ 
\citep[note that $i^{\ast}_{775}=22.3$;][]{goto:05}.}
\end{deluxetable}


\begin{deluxetable}{lrrrrr}
\tablecolumns{5}
\tablewidth{0pc}
\tablecaption{Redshifted $(U-V)_z$:  Mean Color Offset and Scatter\tablenotemark{a}\label{uvztab}}
\tablehead{
  \colhead{Galaxy Type\tablenotemark{b}} & \colhead{$N_g$}  & 
  \colhead{Offset\tablenotemark{c}} & 
  \colhead{$\sigma_{obs}$\tablenotemark{c}} &
  \colhead{$\sigma_{int}$\tablenotemark{d}} }
\startdata
Absorption      & 72 &  0.000$\pm$0.011 &  0.091$\pm$0.009 &  0.083$\pm$0.012\\
E+A             & 17 & -0.192$\pm$0.084 &  0.282$\pm$0.041 &  0.266$\pm$0.044\\
Emission        & 26 & -0.779$\pm$0.129 &  0.564$\pm$0.076 &  0.554$\pm$0.077\\
\\
E+S0            & 94 &  0.005$\pm$0.012 &  0.116$\pm$0.019 &  0.110$\pm$0.020\\
Elliptical Only & 55 & -0.006$\pm$0.015 &  0.095$\pm$0.013 &  0.087$\pm$0.015\\
S0 Only         & 39 &  0.010$\pm$0.043 &  0.197$\pm$0.062 &  0.187$\pm$0.063\\
Spiral+Irregular & 26 & -0.742$\pm$0.109 & 0.497$\pm$0.072 &  0.485$\pm$0.073\\
\enddata
\tablenotetext{a}{The CM relation was determined using the same 67 red 
absorption-line members as the CM relation used in Table~\ref{colortab}.}
\tablenotetext{b}{Spectral types were assigned using [OII]$\lambda3727$, 
H$\delta$, and H$\gamma$ (see \S\ref{spectypes}).  Morphological types were 
classified using HST/ACS imaging \citep[see \S\ref{acs};][]{postman:05}.}
\tablenotetext{c}{The mean offset and observed scatter in $(U-V)_z$ 
as well as the mean colors were determined using the biweight estimator 
\citep{beers:90}.}
\tablenotetext{d}{Following B06, we correct the observed scatter in color 
for measurement errors.}
\end{deluxetable}

\begin{deluxetable}{lrrrrrr}
\tablecolumns{7}
\tablewidth{0pc}
\tablecaption{Indices of Composite Spectra\label{indices}}
\tablehead{
  \colhead{Galaxy Type} & \colhead{$N_g$\tablenotemark{a}}  & 
  \colhead{[OII]\tablenotemark{bc}} & 
  \colhead{H$\delta$\tablenotemark{bc}} & 
  \colhead{H$\gamma$\tablenotemark{bc}} & 
  \colhead{H$\delta_A$\tablenotemark{cd}} &
  \colhead{$D_N(4000)$\tablenotemark{cd}}  }
\startdata
Absorption & 71  &   0.7$\pm$0.0 &   1.7$\pm$0.0 &   0.2$\pm$0.1 &   0.5$\pm$0.0 &   1.67$\pm$0.00   \\
~~~Abs,~~~$0.0\leq\Delta(V-i_{775})\leq0.1$  & 36 &   0.2$\pm$0.1 &   1.3$\pm$0.1 &   0.3$\pm$0.1 &   0.1$\pm$0.1 &   1.73$\pm$0.00   \\
~~~Abs,   $-0.1\leq\Delta(V-i_{775})<0.0$  & 25 &   1.0$\pm$0.1 &   2.1$\pm$0.1 &   1.2$\pm$0.1 &   1.0$\pm$0.1 &   1.66$\pm$0.01   \\ 
E+A	   & 17  &  -0.4$\pm$0.2 &   5.7$\pm$0.2 &   5.9$\pm$0.1 &   4.4$\pm$0.2 &   1.55$\pm$0.01   \\ 
Emission   & 26  & -24.6$\pm$0.6 &   3.8$\pm$0.1 &  -4.4$\pm$0.8 &   4.2$\pm$0.2 &   1.24$\pm$0.01   \\
\\
E+S0	   & 90  &  -0.9$\pm$0.1 &   2.5$\pm$0.0 &   0.8$\pm$0.1 &   1.4$\pm$0.0 &   1.62$\pm$0.00   \\
Elliptical & 53  &  -0.1$\pm$0.1 &   1.9$\pm$0.1 &   0.8$\pm$0.1 &   0.7$\pm$0.1 &   1.66$\pm$0.00   \\
~~~Elliptical:  Red, Bright\tablenotemark{e} & 28 &  -0.0$\pm$0.1 &   2.2$\pm$0.1 &   0.7$\pm$0.1 &   1.3$\pm$0.1 &   1.70$\pm$0.01   \\
S0	   & 37  &  -2.2$\pm$0.3 &   3.3$\pm$0.1 &   0.5$\pm$0.2 &   2.5$\pm$0.1 &   1.58$\pm$0.01   \\
~~~S0: Red, Bright\tablenotemark{e} & 12 &   0.4$\pm$0.1 &   3.1$\pm$0.2 &   2.4$\pm$0.2 &   2.5$\pm$0.2 &   1.67$\pm$0.02   \\ 
Sp+Irr	   & 25  & -23.1$\pm$1.1 &   3.6$\pm$0.2 &  -1.8$\pm$1.1 &   3.8$\pm$0.2 &   1.27$\pm$0.01   \\
Red Mergers\tablenotemark{f} & 17 &   0.9$\pm$0.1 &   2.5$\pm$0.2 &   0.4$\pm$0.6 &   0.5$\pm$0.2 &   1.70$\pm$0.01   \\ 
\\
Field E+A\tablenotemark{g}  &  6 &  -1.5$\pm$0.4 &   4.6$\pm$0.2 &   4.9$\pm$0.3 &   4.3$\pm$0.2 &   1.36$\pm$0.02   \\
\enddata
\tablenotetext{a}{The number of individual spectra combined to determine
the composite spectrum; the five cluster galaxies that do not have 
measured [OII]$\lambda3727$ are excluded.  }
\tablenotetext{b}{We use the same bandpasses defined in \citet{fisher:98} 
to measure these spectral indices; the equivalent widths are in \AA.  
The continuum sidebands are further way
from the line center than those used for measuring, $e.g.$ H$\delta_A$, 
to accommodate the wider wings of strong Balmer absorption.}
\tablenotetext{c}{The errors were determined by generating 1000 composite spectra
from the base set for each galaxy type by bootstrap, measuring the same spectral line, taking 
the biweight scale of the distribution \citep{beers:90}, and dividing that scale by $\sqrt{N_g}$.}
\tablenotetext{d}{To compare to the star formation histories in
  \citet{kauffmann:03}, we use their defined bandpasses to measure H$\delta_A$ (\AA)
  and $D_N(4000)$.}
\tablenotetext{e}{As in Table~\ref{colortab}, members are bright and red if 
 $i_{775}<22$ and $\Delta(V-i_{775})\geq-0.2$.}
\tablenotetext{f}{Composite spectrum of the 17 members that are spectroscopically confirmed 
to be in red, merging systems by \citet{tran:05a}}
\tablenotetext{g}{Composite field E+A spectrum ($0.3<z<1$; $\bar{z}=0.6$) from \citet{tran:04a}.}
\end{deluxetable}

\clearpage

\end{document}